\documentclass[preprint]{revtex4}  
\usepackage{amsmath}
\usepackage{amssymb}
\usepackage{amsbsy}
\usepackage{graphicx}
\usepackage{wasysym}
\usepackage[dvips]{epsfig}
\usepackage{psfrag}
\usepackage[ansinew]{inputenc}
\usepackage[T1]{fontenc}
\usepackage{lmodern}
\usepackage{color}

\newcommand{\ds}{\displaystyle}
\newcommand{\ra}{\rightarrow}
\newcommand{\sca}{\!\cdot\!}

\newcommand{\p}{\partial}
\newcommand{\n}{\nabla}

\newcommand{\epsi}{\varepsilon}

\newcommand{\bra}[1]{\left\langle{#1}\right|}

\newcommand{\norm}[1]{\left|\left| #1 \right|\right|}
\newcommand{\op}[1]{\mathsf #1}
\newcommand{\set}[1]{\mathcal{#1}}

\renewcommand{\vec}[1]{\mbox{\boldmath $ #1 $}}
\newcommand{\R}{\mathbb{R}}

\def\gu{\;{\lower0.3ex\hbox{$\buildrel > \over {\scriptstyle \sim}$}}\;}
\def\lu{\;{\lower0.3ex\hbox{$\buildrel < \over {\scriptstyle \sim}$}}\;}

\def\Xint#1{\mathchoice
   {\XXint\displaystyle\textstyle{#1}}%
   {\XXint\textstyle\scriptstyle{#1}}%
   {\XXint\scriptstyle\scriptscriptstyle{#1}}%
   {\XXint\scriptscriptstyle\scriptscriptstyle{#1}}%
   \!\int}
\def\XXint#1#2#3{{\setbox0=\hbox{$#1{#2#3}{\int}$}
     \vcenter{\hbox{$#2#3$}}\kern-.5\wd0}}

\def\dashint{\Xint-}

\begin{document}

\title{Active plasma resonance spectroscopy:\\ A kinetic functional analytic description}
\author{J. Oberrath}
\author{R.P. Brinkmann}
\affiliation{Institute for Theoretical Electrical Engineering,\\  
             Department for Electrical Engineering and Information Technologies,\\ Ruhr University Bochum, D-44780 Bochum, Germany}
\date{\today}
\begin{abstract}

The term \textit{Active Plasma Resonance Spectroscopy} (APRS) denotes a class of related techniques which utilize, for~diagnostic purposes, 
the natural ability of plasmas to resonate on or near the electron plasma frequency $\omega_{\rm pe}$:  
A radio frequent signal (in the GHz range) is coupled into the plasma via an antenna or probe, 
the~spectral response is recorded,  and a mathematical model 
is used to determine plasma parameters like the electron density or the electron temperature. \linebreak
This manuscript  provides a kinetic description of APRS valid for all pressures and probe geometries. 
Subject of the description is the interaction of the probe with the plasma of its influence domain. \linebreak
In~a first step, the kinetic free energy of that domain is established which has a definite time derivative with respect
to the RF power. In the absence of RF excitation, it assumes the properties  of a Lyapunov functional; 
its minimum provides the stable equilibrium of the plasma-probe system.  \linebreak
Equipped with a scalar product motivated by the second variation of the free energy, the set of all 
perturbations of the equilibrium  forms a Hilbert space. The dynamics of the perturbations can be 
cast in an evolution equation in that space.  The spectral response function of the plasma-probe system 
consists of matrix elements of the resolvent of the dynamical operator.
An interpretation in  terms of an equivalent electric circuit model is given and the residual broadening of the spectrum  
in the collisionless regime is explained.

\end{abstract}
\maketitle

\section{Introduction}

\begin{figure}[h!]
\includegraphics[width= 0.8\columnwidth]{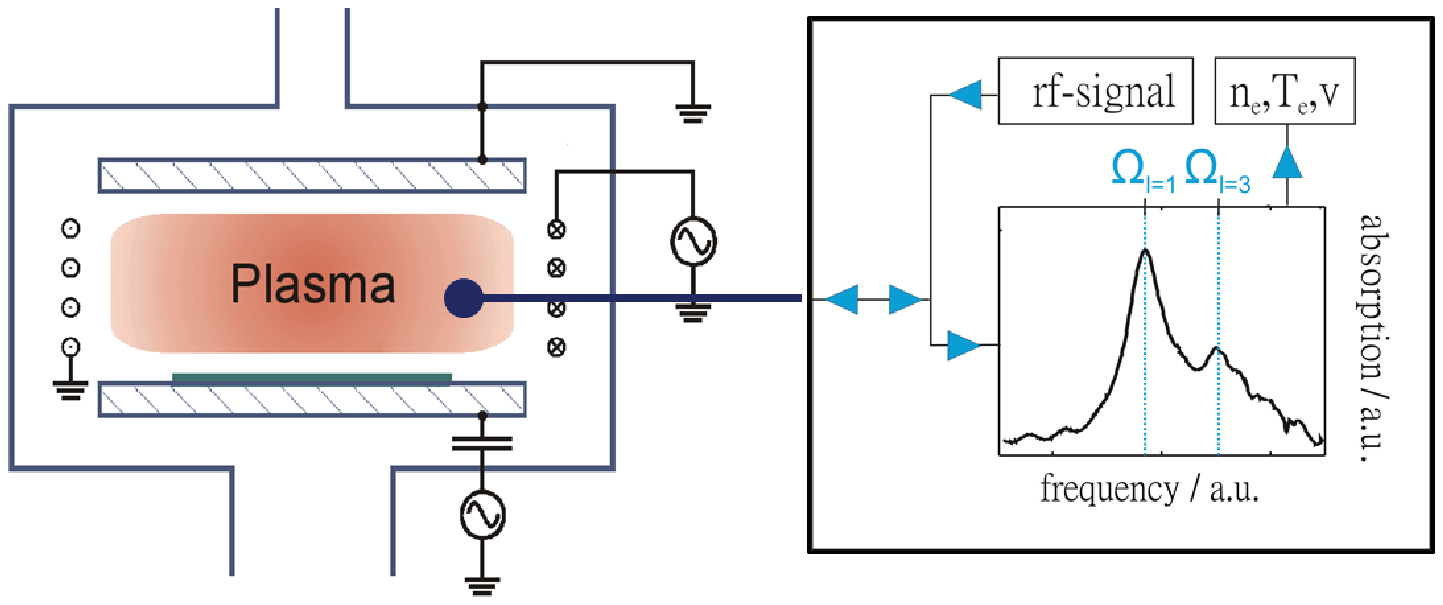}
\caption{Schematic of active plasma resonance spectroscopy: A radio frequent signal (GHz range) is coupled into the plasma via a electric probe, 
         the spectral response is recorded, and a mathematical model is used to determine parameters like the electron density or the electron temperature.
}\label{APRS}
\end{figure}

The term \textit{Active Plasma Pesonance Spectroscopy} (APRS) denotes a wide class of related techniques which utilize, for~diagnostic purposes, 
the natural ability of plasmas to resonate on or near the electron plasma frequency $\omega_{\rm pe}$. 
The basic idea dates back to the early days \linebreak of discharge physics \cite{TonksLangmuir1929,Tonks1931}
but has recently found renewed interest  as an approach to
industry-compatible plasma diagnostics: A radio frequent signal in the GHz range is coupled into the plasma via an electric probe (see fig.~\ref{APRS}), 
the~spectral response is recorded (with the same\linebreak  or with another probe),  and a mathematical model 
is used to determine plasma para\-meters like the electron density $n_{\rm e}$ or the electron temperature $T_{\rm e}$.  
Compared with other plasma diagnostics techniques, for example Langmuir probe analysis, APRS 
has many advantages.
\linebreak
Particularly important for industrial application is its insensitivity against contamination; 
this feature makes APRS ideal for the diagnostics and supervision of plasma-assisted deposition of dielectrics and similar manufacturing processes.

In the course of the last fifty years, many variants of APRS have been proposed \cite{stenzel1976, kim2003, piejak2004, wang2011, takayama1960, messiaen1966, waletzko1967, vernet1975, sugai1999, blackwell2005, scharwitz2009, lapke2008}. \linebreak
According to ref.~\cite{lapke2011}, they may be classified as follows: \textit{Electromagnetic} methods \cite{stenzel1976, kim2003, piejak2004, wang2011} excite 
cavity or transmission line resonances which are  already present under vacuum conditions. \linebreak
In the presence of plasma these resonances are shifted, and a qualitative analysis -- based on the dispersion relation  $\omega^2 = c^2 k^2 + \omega_{\rm pe}^2$ 
of an electromagnetic wave in a homogeneous plasma -- \linebreak   predicts that the shift of the squared frequency 
$\Delta \omega^2 = \omega^2-\omega_0^2$ 
is proportional to the local electron density $n_{\rm e}$.
\textit{Electrostatic} techniques \cite{takayama1960, messiaen1966, waletzko1967, vernet1975, sugai1999, blackwell2005, scharwitz2009, lapke2008}, in contrast, 
excite surface wave modes which vanish at zero plasma density. 
In this case, the dispersion relation $\omega^2 = s\, k\, \omega_{\rm pe}^2$  of a long electro\-static surface wave 
propagating along a homogeneous plasma boundary sheath of thickness $s$
before a conductor suggests that the squared resonance frequency $\omega^2$\linebreak  itself
is proportional to the local electron density.

To be useful for quantitative diagnostics, however, the analysis must proceed beyond such qualitative arguments.
The exact nature of the spatial average implied in the term ''local plasma density'' must be specified, 
and the constant of 
proportionality between that quantity and the shift or absolute value  of the squared resonance frequency must be found. \linebreak
In other words: Mathematical models of the probes and their interaction with the surrounding plasma must be established.
Many researchers have made attempts at this task \cite{fejer1964, harp1964, 
crawford1964, dote1965, kostelnicek1968, cohen1971, tarstrup1972,
aso1973, bantin1974, booth2005, walker2006, lapke2007, xu2009,
xu2010, li2010, liang2011,ichikawa1962, buckley1966, balmain1966, hellberg1968, li1970,
meyer1975, nakatani1976}, using analytical and/or numerical techniques and plasma models of different complexety.

\begin{figure}[h!]
\includegraphics[width= 0.8\columnwidth]{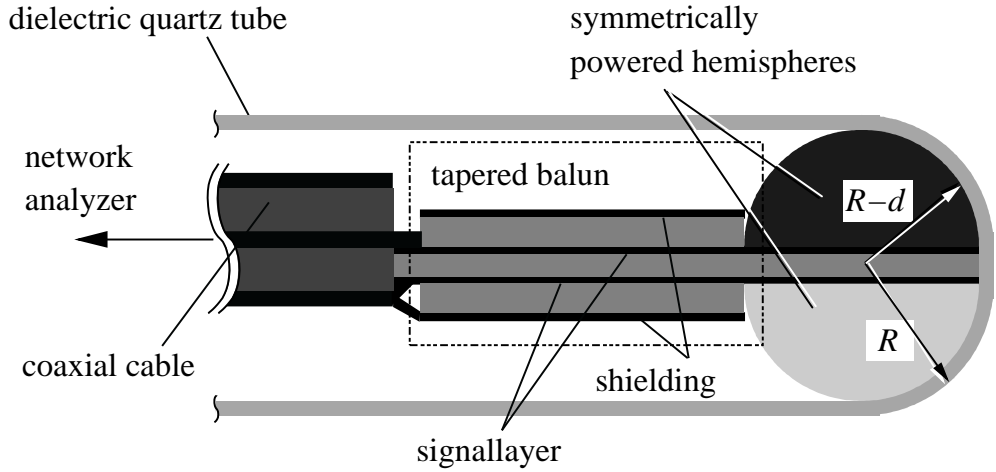}
\caption{Schematic cross section of the multipole resonance probe prototype as an example of an electrostatic probe design \cite{lapke2011} (reprint). The probe itself consists of two metallic hemispheres of radius R-d, symmetrically driven via a tapered balun transformer realized as a 4-layer printed circuit board. The probe system is included in a cylindrical quartz tube of a thickness d.
}\label{mrp}
\end{figure}

The cited publications have in common that they all concentrate on specific probe designs. \linebreak
Members of our group, for example, have analyzed the \textit{Multipole Resonance Probe (MRP)}, \linebreak
an~optimized variant of electrostatic APRS \cite{lapke2008,lapke2011}. (See fig.~\ref{mrp}.) 
 However, it is also of interest to 
study generic features of APRS which are independent of any particular realization. \linebreak
Using methods of functional analysis, members of our group have recently presented such an abstract study of 
electrostatic APRS \cite{lapke2013}. The main result of that investigation was that, for any possible probe design, 
the spectral response function could be expressed as a matrix element of the resolvent of the dynamical operator. 

Unfortunately, the validity of the results in \cite{lapke2013} is limited because the study was based on the cold plasma model,
and could thus not capture the  kinetic effects which are influential in the low pressure regime.
The importance of these effects was emphasized by \cite{lapke2011} which introduced an effective damping 
as $\nu_{\rm eff} = \nu_{\rm m} + \nu_{\rm kin}$, where  $\nu_{\rm m}$ (proportional to the pressure)\linebreak 
is the electron-neutral collision rate, and $\nu_{\rm kin}$ (pressure-independent)  mimics kinetic effects: \linebreak
At $p=10\,{\rm Pa}$, the kinetic collision frequency $\nu_{\rm kin}=3.5\times 10^{8}\,{\rm s}^{-1}$
was found to exceed the momentum transfer rate $\nu_{\rm m}=1.7\times 10^{8}\,{\rm s}^{-1}$ by more than a factor of two.

This manuscript aims to close the gap. We will present a fully kinetic generalization of the study of \cite{lapke2013},
i.e., an abstract kinetic model of electrostatic APRS valid for all pressures. \linebreak
It will turn out that many insights can be directly transfered. In particular, it still holds that, \linebreak for any  possible probe design, 
the spectral response of the probe-plasma system can be expressed as a  matrix  element of the resolvent of the dynamical operator.
The physical content \linebreak  of this expression,  however, will prove very different. (Unfortunately, also the mathematical complexity
of its derivation will be much higher.)

The rest of the manuscript is organized as follows: In section \ref{sec:model} a kinetic model for the interaction of an electrostatic APRS probe 
(of arbitrary geometry) with a low temperature plasma (of arbitrary pressure) is derived. 
 Section \ref{sec:thermo} explores an analogy to classical thermo\-dynamics and defines the kinetic free energy $\mathfrak{F}$, 
which is shown to be a Lyapunov functional. \linebreak 
Its minima correspond to the non-RF excited equilibria of the system of plasma and probe. \linebreak
Section \ref{sec:lin} linearizes the dynamical equations around such an equilibrium and establishes the quadratic free energy $\mathbb{F}$,
a positive definite quadratic functional in the distribution function. In section \ref{sec:func}, this functional is employed 
to define a suitable Hilbert space which then allows to formulate the spectral response of the system as matrix elements of the 
resolvent of the time evolution operator. A summary and conclusions are given in section \ref{sec:conclusion}.

\section{The interaction of an electric probe with a plasma}\label{sec:model}

\begin{figure}[h!]
\includegraphics[width= 0.8\columnwidth]{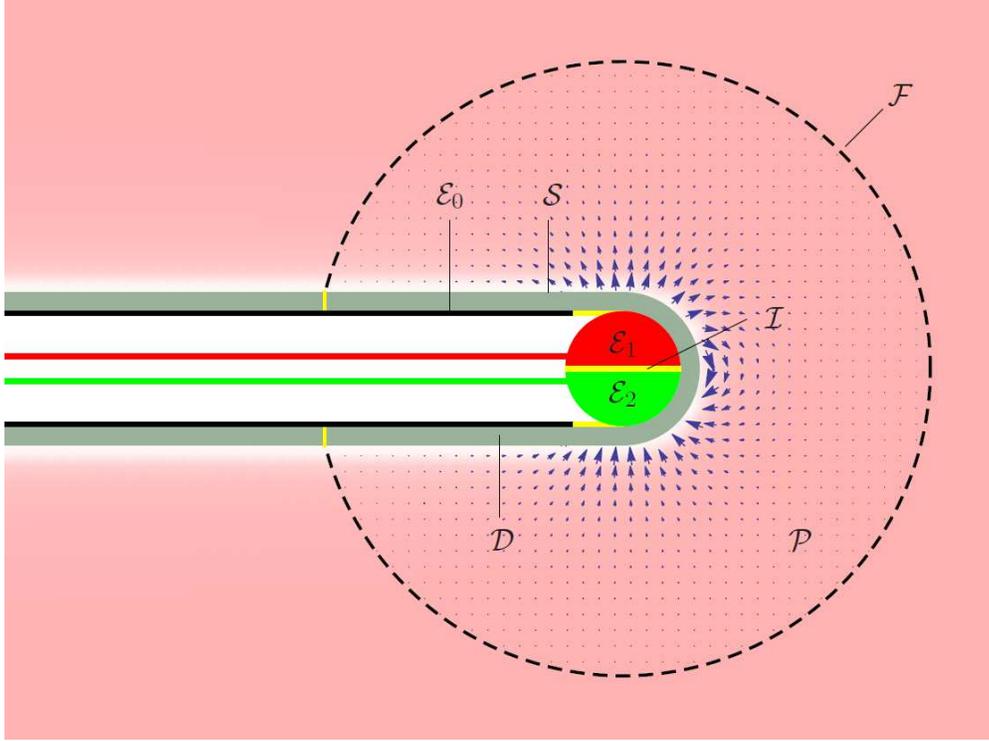}
\caption{Illustration of the abstract model of electrostatic APRS. (The figure is realistic for a multipole resonance probe 
         of a diameter $\diameter=8\,{\rm mm}$ in a plasma with $n_{\rm e} = 10^{10}\,{\rm cm}^{-3}$ and $T_{\rm e}= 3\,{\rm eV}$. 
         The probe consist of N electrodes ${\set E}_n$ -- here only two,  red and green -- connected to the applied RF; 
				 separated from each other and from the grounded shield ${\set E}_0$ (black) by yellow insulators $\set I$ (yellow). 
				 The~whole probe is covered by a dielectric medium $\set D$ (grey) and immersed into a plasma $\set P$ (pink)
				 which developes a sheath (white) close to material surfaces. The outer, plasma-facing surface of the dielectric is
				 denoted by $\set S$. The distant circle indicates the boundary of the influence domain~$\set V$; it consists of the interface $\set F$ (dashed)
				 and some insulator $\set I$ (yellow). Within the influence domain, the applied RF field is indicated by the blue arrows.}\label{abstractmodel}
\end{figure}

As explained in the introduction, we strive in this paper to establish a kinetic model of 
the interaction of an electric probe with a plasma. The considered set-up is depicted in fig.~\ref{abstractmodel}: \linebreak
A discharge is operated in a chamber, and an dielectrically covered electric probe is inserted, fed~by an RF signal via a
shielded cable. 
To render our considerations general, we assume that the 
probe consists of an arbitrarily shaped head with a finite 
number $N$ of electrodes $\set E_n$. \linebreak
Grounded metal surfaces, for example the outer conductor (mantle) of the shielded RF cable, \linebreak are treated as an additional electrode $\set E_0$. 
All electrodes are ideal (infinite conductivity), and are shielded from each other by ideal isolators $\set I$ (zero conductivity and permittivity).
\linebreak
The voltages $U_n$ applied at them constitute the excitation of the system; of course, $U_0 \equiv 0$. The currents $I_n$ represent the response. 
As for the dielectric cover $\set D$, we assume that it has a permittivity $\epsi_{\rm D}(\vec r)$ that is temporally but not 
necessarily spatially constant.

Obviously, it cannot be our goal to describe the system of plasma and probe \textit{completely}: \linebreak
This would be equivalent to a full ab-initio simulation, 
for~which diagnostics algorithms have typically neither 
data nor numerical resources. 
Instead, we aim for a \textit{model of the influence domain of the probe}, 
i.e., of the spatial region $\set V$ which is directly influenced by the probe. \linebreak
(Note that ``influence'' refers to the perturbation by the applied RF. The static reaction of the plasma to the mere presence 
of the probe  as a material object is not contained in our dyn\-amical model but reflected in the
assumed equilibrium of the system of plasma and probe.) \linebreak
Details are shown in fig.~\ref{abstractmodel}: 
The influence domain $\set V$ consists of the dynamically perturbed part of the plasma $\set P$
and of the dielectric probe cover $\set D$. It is bounded by the electrodes $\set E_n$, \linebreak 
the insulator $\set I$, and the interface $\set F$  between the perturbed and the unperturbed plasma. 
(This interface -- of course only a convenient imagination -- will be discussed in detail below.) \linebreak
The plasma-facing surface of the dielectric cover, i.e., the intersection of $\set P$ and $\set D$, is called $\set S$.\linebreak
As for the size of the influence domain $\set V$, 
we assume its length scale $R_{\set V}$ as much smaller than the dimension $L$ 
of the plasma chamber, the energy
diffusion length $\lambda_\epsilon$, and the skin depth $\lambda_{\rm s}$.  \linebreak
(In low pressure discharges at, say, $1-10\,{\rm Pa}$, those length scales are typically comparable.) 
On the other hand, the length scale $R_{\set V}$   is taken to be larger than the elastic mean free path $\lambda$ of the electrons, 
the radius of the probe tip $R$, and the electron Debye length $\lambda_{\rm D}$.
Altogether, the assumptions on the length scales lead to the relative ordering (``regime'') 
$\lambda_{\rm D} \lu \lambda \approx R \ll 	R_{\set V} \ll L \approx \lambda_{\epsilon} \approx \lambda_{\rm s}$.

In general, a low temperature plasma discharge is far from thermodynamic equilibrium. Even when stationary, its equilibrium is \textit{dissipative}; 
i.e.,~characterized by a steady exchange of energy and matter with the environment and by a corresponding production of entropy. 
However, this equilibrium applies to the whole discharge and is established relatively slowly.
When we focus on the described small influence domain $\set V$, and  only on a comparatively fast time scale -- 
namely that  of electro\-static plasma oscillations --, we can neglect all
dissipative phenomena and treat the plasma as being in a \textit{nondissipative} equilibrium.
This means that the electron distribution function is a scalar function of the total energy,
\begin{align}
{\bar f}(\vec r,\vec v) = \bar{F}\left(\frac{1}{2}m\vec v^{2}-e\bar{\Phi}(\vec r)\right).
\label{Equilibrium1}
\end{align}
The equilibrium potential $\bar{\Phi}$ must fulfill Poisson's equation, with $n_{\rm i}(\vec r)$ 
as the net ion density and $\epsi(\vec{r})=\epsi_0\epsi_{\rm r}(\vec{r})$ as the permittivity
 ($\epsi_{\rm r}$ is $1$ in $\set P$ and equal to $\epsi_{\rm D}(\vec r)$ in $\set D$): 
\begin{equation}
 -\nabla\!\cdot\!\left(\epsi(\vec{r})\n{\bar\Phi}\right) 
= \left\{\ 
\begin{matrix} 
0 & & \vec r & \in & \set D\\[0.5ex] 
\ds e n_{\rm i}(\vec r)-e \int_0^\infty  \bar F\left(\frac{1}{2}m v^{2}-e\bar{\Phi}(\vec r)\right) 4\pi v^2 dv & & \vec r & \in & \set P 
\end{matrix}
  \right. \ .
  \label{Equilibrium2}
\end{equation}
It must also obey the static boundary conditions that it is zero at the electrodes, $\bar\Phi=0$ at $\set E_n$, \linebreak
and has a vanishing normal at the insulators, $-\vec n\!\cdot\!\nabla\bar\Phi=0$ at $\set I$. 
In addition to the charges of the electrons and ions, also the surface charges $\sigma(\vec{r})$ at the plasma-facing dielectrics must be considered, 
namely by the transition condition
$\left.\left(\epsi_{\rm D}(\vec r)\n\bar\Phi^{(\set D)}-\epsi_0\n \bar\Phi^{(\set P)}\right)\!\cdot\!\vec n\right|_{\set S}=\sigma(\vec{r})$. 

Note that these equations do not establish a complete description of the equilibrium state:
The ion density $n_{\rm i}(\vec r)$ is not specified; this would require information on the ion dynamics which is 
not available to the model. Likewise, the surface charge density $\sigma(\vec{r})$ is not specified;
this would require not only information on the ion dynamics but also on the electron sticking factor of the surface.
Finally, the actual form of the energy distribution $\bar{F}(\epsilon)$ is not specified,
this would require information on the dissipative dynamics of the plasma, i.e., on the heating and
cooling mechanisms. We can, however, assume that $\bar{F}$ is physical and stable in terms of fast electrostatic modes;
i.e., positive ($\bar{F}\left(\epsilon\right)> 0$), 
monotonically decreasing $\left(\frac{\partial \bar{F}}{\partial \epsilon}<0\right)$, \linebreak 
and faster vanishing than any power $\left(\lim_{\epsilon\ra\infty} {\epsilon}^{\,n} \bar{F}\left({\epsilon}\,\right)=0\ ,\ n\geq0\right)$. 
We call such a function a \textit{qualitative exponential}.  
Its inverse $\bar{\epsilon}(F)$ exists and is called a \textit{qualitative logarithm}; \linebreak
it is defined on the interval $[0, \bar{F}_{\rm max}]$ and can be expanded to the 
complete positive axis with the properties $\frac{\partial\bar{\epsilon}}{\partial F}< 0$, $\lim_{F\ra 0} \bar{\epsilon}(F)=\infty$, 
and $\lim_{F \ra \infty} \bar{\epsilon}(F)  =-\infty$. In the special case \linebreak of a Maxwellian distribution,
$\bar{F}_{\rm M}(\epsilon) = \hat{F} \exp(-\epsilon/T_{\rm e})$ and $\bar\epsilon_{\rm M}(F) = -T_{\rm e} \ln(F/\hat{F})$. 

The assumed equilibrium is perturbed by the measurement process, and it is the reaction \linebreak of the system that we aim to describe. 
The dynamical variable of the model is the electron distribution function $f(\vec r,\vec v,t)$ whose temporal evolution 
is governed by a kinetic equation. \linebreak 
As for the time scale, we focus on the  RF frequency $\omega_{\rm RF}$, assumed to be smaller but similar to the
electron plasma frequency $\omega_{\rm pe}$, possibly comparable to the elastic collision frequency $\nu_0$ \linebreak
and much larger than the inelastic collision frequency $\nu_{\rm i}$, the ion plasma frequency $\omega_{\rm pi}$, 
and~the slow frequencies $\omega_{\rm g}$ of all neutral gas phenomena, 
$\omega_{\rm pe} \gu \omega_{\rm RF}\gu \nu_0 \gg \nu_{\rm i} \approx  \omega_{\rm pi} \gg \omega_{\rm g}$. \linebreak
In accordance with our assumptions discussed above, we focus on non-dissipative processes. 
Inelastic and ionizing electron-neutral collisions and Coulomb interaction between free charge carriers are  neglected.
Elastic electron-neutral collisions are treated in the limit $m_{\rm e}/m_{\rm N}\to 0$; \linebreak
i.e., the neutrals are seen as immobile scattering centers with a velocity and angle dependent \linebreak
differential collision frequency $\frac{d\nu}{d\Omega}(v,\vartheta)$. 
The total collision frequency is
$\nu_0(v) = \int_\Omega \frac{d\nu}{d\Omega} \,d\Omega$. \linebreak
(Here, $v=|\vec{v}|$, $\hat{\vec{v}}= \vec{v}/v$, $\cos(\vartheta) = \hat{\vec{v}}\cdot\vec{e}$; $\vec{e}$ is the unit vector related 
to the differential $d\Omega$.) \linebreak
Thus, we describe the electron dynamics in the plasma domain $\set P$ by a reduced version of the 
kinetic (or Boltzmann) equation  
 \begin{equation}
 \frac{\p f}{\p t}+\vec{v}\!\cdot\!\nabla f-\frac{e}{m}\vec{E}\!\cdot\!\nabla_{v}f  =\langle f\rangle_{\rm el} 
=- \nu_0 f + \int_\Omega\frac{d\nu}{d\Omega}f(|\vec v|\vec e)\, d\Omega.
\label{Boltzmann}
\end{equation}

Boundary conditions for $f$ must be given at all boundaries of $\set P$. Before the boundary $\set S$,\linebreak
i.e., in front of all material surfaces, we assume the presence of a floating boundary sheath. \linebreak
The few energetic electrons which overcome the floating  potential of the sheath are assumed to undergo specular reflection
at $\set S$. This is compatible with the neglect of ionization, \linebreak and the assumption that ion densities and 
surface charges do not vary on the  fast time scale. \linebreak
It is also compatible with the equilibrium distribution.
With $\vec n$ denoting the surface normal, \linebreak we thus assume at the plasma-wall boundary $\set S$:
\begin{align}
f(\vec r, \vec v, t)=f(\vec r,\vec v - 2 \,\vec{n}\!\cdot\!\vec{v} \,\vec{n},t)
\quad ,\quad \vec r\in\set S\ .
\end{align}
 At the interface $\set F$ between the influence domain and the outer plasma, we assume that the distribution
function $f$ is close to the equilibrium distribution $\bar f$. We cannot demand that it is exactly \textit{identical}
-- this would prevent any interaction with the unperturbed plasma. 
\linebreak
Instead, we demand that the difference of the distribution functions is small, in a sense that will be made more explicit below:
\begin{equation}
f(\vec r, \vec v, t) 
= \bar F\left(\frac{1}{2}m\vec v^{2}-e\bar{\Phi}(\vec r)\right) 
+ \delta\!f_{\set F}(\vec r, \vec v, t)\quad ,\quad \vec r\in\set F\ .
\label{stoerf1}
\end{equation}

For the calculation of the field, we can adopt the electrostatic approximation $\vec{E}=-\nabla\Phi$:
The skin effect is negligible ($R_{\set V}\ll \lambda_{\rm s}$) and no electromagnetic waves are emitted
($\omega_{\rm RF} \lu \omega_{\rm pe}$). We thus use Poisson's equation, with~the same ion density as in the equilibrium,
\begin{equation}
 -\nabla\!\cdot\!\left(\epsi(\vec{r})\n\Phi\right) 
= \left\{\ 
\begin{matrix} 
0 & & \vec r & \in & \set D\\[1ex] 
\ds e\left(n_{\rm i}(\vec r)-\int_{\R^3} f\ 
d^3v\right) & & \vec r & \in & \set P 
\end{matrix}
  \right. \ .
  \label{Poisson1}
\end{equation}
The surface charges are also idential to those of the equilibrium,
\begin{align}
	  \left.\left(\epsi_{\rm D}(\vec r)\n\Phi^{(\set D)}-\epsi_0\n \Phi^{(\set P)}\right)\!\cdot\!\vec n\right|_{\set S}=\sigma(\vec{r}).
\end{align}
The potential boundary conditions must now reflect the excitation by the RF voltages $U_n$ applied to the 
electrodes ${\set E}_n$. At the influence domain interface $\set F$, it is assumed to be close to the unperturbed,  time independent equilibrium 
potential $\bar\Phi$; we will call the difference $\delta\Phi_{\set F}$.\linebreak
(More specific statements on $\delta\Phi_{\set F}$ will be made below.)
At the insulator surface $\set I$, the~normal component vanishes. Altogether, we assume
\begin{align}
	&\Phi
= \left\{\ 
\begin{matrix} 
U_n & & \vec r & \in & {\set E}_n\ ,\ n\in[0,N]\\[1ex] 
\ds {\bar\Phi}\left(\vec r\right) + \delta\Phi_{\set F}\left(\vec r\right)  & & \vec r & \in & \set F 
\end{matrix}
  \right. \ , \\
 &  \vec n\!\cdot\!\nabla \Phi = 0, \hspace{2.5 cm} \vec r  \in  \set I   \nonumber.
\end{align}

The response of the system to the RF excitation is given by the divergence-free 
current $\vec J$,\linebreak 
the sum of the electron current (in the plasma) and the displacement current:
\begin{equation}
 \vec J(\vec r, t) 
= \left\{\ 
\begin{matrix} 
\ds - e\int_{\R^3} \vec v f\,d^3 v -\epsi_0\frac{\p\n\Phi}{\p t} & & \vec r & \in & \set P \\[1ex] 
\ds -\epsi_{\rm D}(\vec{r})\frac{\p\n\Phi}{\p t} & & \vec r & \in & \set D
\end{matrix}
  \right.\ .
  \label{totStrom}
\end{equation}
At the boundaries ${\set E}_n$ and $\set F$, we count currents as positive when they flow into the plasma.
(At the insulators $\set I$, the normal current density vanishes.)
Considering that the electrodes are dielectrically shielded, we define  
\begin{align}
	&I_n = - \int_{\set E_n} \vec{J}  \sca d^2\vec{r} =  \int_{\set E_n}\epsi_{\rm D}(\vec{r})\frac{\p\n\Phi}{\p t} \cdot d^2 \vec r,\hspace{1.0cm} n=[0,N],  \\[0.5ex]	
	&I_{\set F} =-\int_{\set F} \vec J  \sca d^2\vec{r}  =  \int_{\set F} \left(e\int_{\R^3} \vec v f\,d^3 v +\epsi_0\frac{\p\n\Phi}{\p t}\right) \cdot d^2\vec{r}.	\label{Kirchhoff}
\end{align}
For these currents, Kirchhoff's law can be established. 
The total current is divergence-free, \linebreak hence its surface integral over the boundary $\partial\set V$ vanishes, and we obtain
\begin{align}
 \sum_{n=0}^N I_n+ I_{\set F} \equiv -\int_{\partial\set P} \vec J  \sca d^2\vec{r}  = 0.	\label{Kirchhoff}
\end{align}


\section{The ``thermodynamics'' of the plasma-probe system}\label{sec:thermo}

In the last section, we have formulated a model for the subdomain $\set V$, i.e., for the probe and its immediate 
vicinity which is in close contact with the unperturbed rest of the plasma. \linebreak
In classical thermodynamics, analogous situations are studied where a system with internal energy $U$ and entropy $S$ 
exchanges mechanical work at a rate $P$ and heat at a rate $\dot Q$ with the \linebreak environment of a given temperature $T$.
The first law of thermodynamics states $\frac{dU}{dt} = P + \dot Q$;
the increase of the internal energy $U$ equals the net energy obtained from the environment. \linebreak 
The second law states $\frac{dS}{dt} \ge \frac{1}{T} \dot Q$; the increase of the entropy $S$ is larger than the
influx of the equilibrium entropy. (In the limit case of a reversible processes, the two quantities are equal.) 
Both laws can be combined to demonstrate that  the (Helmholtz) free energy $F=U-TS$ has a definite time derivative with respect to the power, 
$\frac{dF}{dt}\le  P$. In a nonequilibrium plasma,
of~course, classical thermodynamics does not apply, but we will demonstrate in this chapter that a similar construction is 
possible nonetheless: We will define a kinetic equivalent of $F$,  the \textit{kinetic free energy} $\mathfrak F$, 
as the difference of the total energy $\mathfrak U$ and the kinetic entropy $\mathfrak S$. \linebreak
This quantity will also have a defined time derivative in comparison to the (electrical) power,
provided that the dynamics is restricted to the\textit{ nondissipative} processes included in \eqref{Boltzmann}.\linebreak
(In another context -- stability analysis --, the concept was employed, e.g., by \cite{fowler1963,morrison1989,batt1995,spatschek1990}.) 

To derive an expression for $\mathfrak U$, we first integrate \eqref{Boltzmann} over $\set P$ with weight 
$\frac{1}{2}m {\vec v}^2$ to obtain the balance of the kinetic electron energy. Employing Gauss' law and making use 
of the boundary conditions for $f$ at the surface  $\set S$ and the interface $\set F$, we arrive at the following,
where the term on the left represents 
the Ohmic heating of the electrons by the electric field; there are no losses as inelastic collisions were neglected:
\begin{equation}
 \frac{d}{dt}\int_{\set P}\int_{\R^3} \frac{1}{2}m {\vec v}^2 f d^3v\,d^3r + \int_{\set F} \int_{\R^3} \frac{1}{2}m {\vec v}^2 \vec{v} \,\delta\!f_{\set F} d^3v\sca d^2\vec r
=\int_{\set P} e\n\Phi\!\cdot\!\int_{\R^3} \vec{v} f d^3v\, d^3 r.
\label{kinenergybalance}
\end{equation}
Further, we integrate $\nabla\sca\vec{J}=0$ over $\set V$ with weight $\Phi$  for the balance of the field energy: 
\begin{equation}
\frac{d}{d t}\int_{\set V}\frac{1}{2}\epsi \n\Phi^2 \, d^3r- \sum_{n=1}^N U_n I_n + \int_{\set F} ({\bar\Phi}\left(\vec r\right) + {\delta\Phi}_{\set F}\left(\vec r\right))\vec J \sca d^2\vec r = 
-\int_{\set P} e\n\Phi\!\cdot\!\int_{\R^3} \vec{v} f d^3v\, d^3 r.
\label{elstaenergybalance}
\end{equation}
The sum of the two equations yields the balance for the total energy $\mathfrak{U}$,
\begin{align}
 \frac{d}{d t}\left(\int_{\set V}\frac{1}{2}\epsi \n\Phi^2 \, d^3r\right. &+ 
 \left.\int_{\set P}\int_{\R^3} \frac{1}{2}m {\vec v}^2 f d^3v\,d^3r \right)\\[1ex]
&+ \int_{\set F}\left( ({\bar\Phi}\left(\vec r\right) + {\delta\Phi}_{\set F}\left(\vec r\right))\vec J 
 + \int_{\R^3} \frac{1}{2}m {\vec v}^2 \vec{v}\, \delta\!f_{\set F}\, d^3v\right)\!\cdot\! d^2\vec r
= \sum_{n=1}^N U_n I_n.\nonumber
\label{totalenergybalance}
\end{align}

Next, consider the balance of the ``kinetic entropy density'' $\sigma$ and the associated flux ${\vec\Gamma}_\sigma$. \linebreak
These quantities are defined similarly to their thermodynamic analogues, except that not the \linebreak exact logarithm $\ln(f)$ is used, 
but the ``qualitative logarithm'' $\bar\epsilon(f)$. 
(Recollect that the direct thermodynamic analogy of $\bar\epsilon(f)$  
is $-T_{\rm e} \ln(f/\hat{F})$, this explains the ``missing'' factor $T_{\rm e}$): 
\begin{align}
&\sigma  =  \int_{\R^3} \int_0^f \bar\epsilon(f')\, df'\, d^3 v,\\[1ex]
&{\vec\Gamma}_\sigma  =  \int_{\R^3} {\vec v}\int_0^f \bar\epsilon(f')\, df'\, d^3 v.
\label{entropyflux}
\end{align}

Multiplying the kinetic equation \eqref{Boltzmann} with $\bar\epsilon(f)$ and integrating over velocity space yields the kinetic 
local entropy balance, a special case of Boltzmann's H-Theorem \cite{boltzmann1896}:  
\begin{equation}
\frac{\p\sigma}{\p t} + \n\sca{\vec\Gamma}_\sigma =
	\int_{\R^3} \bar\epsilon(f)\left(-\frac{e}{m}\n\Phi\sca\n_v f+\langle f\rangle_{\rm el}\right)\, d^3 v \ge 0.
	\label{localentropy}
\end{equation} 
For a proof of the inequality, we use $\bar\epsilon(f) \n\Phi\cdot \n_v f  = 
 \n_v\cdot\left(\n\Phi \int_0^f \bar\epsilon(f') \, df'\right)$ on the first term on the right 
and employ Gauss' theorem.
This shows that the term is zero, as $f\to 0$ for $|\vec v|\ra\infty$ \linebreak
and $\int_0^f \bar\epsilon(f') \, df'\to 0$ for $f\to 0$, sufficently fast. 
For the second term, we use 
 $\nu_0 = \int_\Omega \frac{d\nu}{d\Omega} \,d\Omega^\prime$, 
split into a sum of two identical terms divided by two, and exchange 
$\vec e \leftrightarrow \vec e^\prime$ in one of them. 
The result is non-negative because $\bar\epsilon(f)$ is monotonically decreasing in $f$.
Note that equality only holds when $f(v{\vec e})=f(v\vec e')$, i.e., when $f$ is isotropic: 
\begin{align}
  \int_{\R^3} \bar\epsilon(f)\langle f \rangle_{\rm el}\, d^3v 
 &=\int_0^\infty v^2\int_\Omega \bar\epsilon(f)
  \left(-\nu_0 f(v{\vec e}) + 
  \int_{\Omega '}\frac{d\nu}{d\Omega}f(v{\vec e'})\,d\Omega '\right)\,d\Omega\,dv\label{HTheorem}\\[0.5ex] 
&=\int_0^\infty \frac{v^2}{2}\int_\Omega\int_{\Omega '}\frac{d\nu}{d\Omega}   
  \Bigl(\bar\epsilon(f(v{\vec e}))-\bar\epsilon(f(v \vec e'))\Bigr)
  \Bigl(-f(v{\vec e})+f(v\vec e')\Bigr)\,d\Omega '\,d\Omega\,dv \ge 0.\nonumber
\end{align}
Now we integrate \eqref{localentropy} over $\set P$ to obtain a balance for the kinetic entropy $\mathfrak{S}$. Note that the 
surface integral over $\set S$ vanishes due to the boundary conditions on $f$:
\begin{equation}
\frac{d}{dt} \int_{\set P}\int_{\R^3} \int_0^f \bar\epsilon(f')\, df'\, d^3 v \,  d^3r  + \int_{\set F} {\vec\Gamma}_\sigma \sca d^2\vec r \ge 0.
\end{equation}
On $\set F$, the distribution $f$ is close to $\bar{f}$.
 We can thus approximate  
\begin{eqnarray}
\int_0^f \bar\epsilon(f')\,df' 
 \approx    \int_0^{\bar{f}} \bar\epsilon(f')\,df' 
     +      \left(\frac{1}{2}m\vec{v}^2-e\bar{\Phi}(\vec r)\right)\delta\!f_{\set F} + \frac{1}{2} \epsilon'(\bar{f})\delta\!f_{\set F}^2.
\label{entropyapprox}     
\end{eqnarray}
Introducing \eqref{entropyapprox} in the definition \eqref{entropyflux} yields the important fact that the 
entropy flux on $\set F$ is,\linebreak in first order approximation, identical to the energy flux on $\set F$. 
Terming the difference $\Delta\vec\Gamma_{\sigma\set{F}}$,\linebreak we write the entropy flux at the boundary  
\begin{equation}
\vec\Gamma_\sigma\bigl\vert_{\set F}
\approx
\int_{\R^3}  \left(\frac{1}{2}m\vec{v}^2-e\bar{\Phi}(\vec r)\right)\vec v f d^3v  + o\!\left((f-\bar{f})^2\right)
= \int_{\R^3} \frac{1}{2}m\vec{v}^2 \vec vf d^3v + \Phi \vec J + \Delta\vec\Gamma_{\sigma\set{F}}.
\label{entropyfluxapprox}
\end{equation}

We are thus moved to define the kinetic free energy ${\mathfrak F}$ as the difference of the total energy $\mathfrak{U}$\linebreak 
 and the kinetic entropy $\mathfrak{S}$. It is interpreted as a functional of the distribution 
function $f$, as~the potential $\Phi$ is determined by Poisson's equation and thus directly coupled to $f$: 
\begin{equation}
 {\mathfrak F}\{f\} = {\mathfrak U}\{f\}-{\mathfrak S}\{f\}  =\int_{\set V} \frac{1}{2}\epsi\n\Phi\{f\}^2 \, d^3r + 
 \int_{\set P} \int_{\R^3}\left(\frac{1}{2}m\vec{v}^2 f 
-\int_0^f \bar\epsilon(f')\,df'\right) d^3v\, d^3 r.
\label{KineticFreeEnergy}
\end{equation}
The time derivative of the kinetic free energy fulfills the inequalities
\begin{equation}
\frac{d{\mathfrak F}}{dt} \le \sum_{n=1}^N U_n I_n  -  \int_{\set F}  \Delta\vec\Gamma_{\sigma\set{F}}  \sca d^2\vec{r} \le \sum_{n=1}^N U_n I_n,
\label{KineticFreeEnergyInequality}
\end{equation}
where the second inequality holds under the \textit{physical assumption} that the excess entropy flux \linebreak 
through the interface $\set F$ (identified as a net free energy loss)  has a definite sign, 
\begin{align}
  \int_{\set F} \Delta\vec\Gamma_{\sigma\set{F}} \sca d^2\vec{r} = - \Delta\dot{\mathfrak F}_{\sigma\set{F}}	  \ge 0.
\end{align}

Relation \eqref{KineticFreeEnergy} and \eqref{KineticFreeEnergyInequality} show that it is indeed possible to define a 
``kinetic free energy'' $\mathfrak{F}$ \linebreak as a direct analogon to the thermodynamic free energy $F$.
The established quantity has many important properties. For example, following refs.~\cite{fowler1963,morrison1989,batt1995,spatschek1990}), 
it can be used to demonstrate that the assumptions on the equilibrium and the dynamics are compatible: 

Consider the case of zero RF excitation of the probe. The right side of \eqref{KineticFreeEnergyInequality} then vanishes, \linebreak
and the kinetic free energy monotonically decreases. According to the theory of Lyapunov, 
the stable equilibria of the system are the minima of the functional. Call such a minimum $f^*$.\linebreak
Around $f^*$, the first variation of the functional ${\mathfrak F}$ has to vanish:
\begin{equation}
 \delta {\mathfrak F} 
=\int_{\set V}\left(\epsi\n\Phi^*\cdot\n\delta\Phi
+\int_{\R^3}\left(\frac{1}{2}m\vec{v}^2 \delta f 
-\bar\epsilon(f^*)\delta f \,\right)\, d^3v\right)\,d^3 r  \stackrel{!}{=} 0.
\end{equation}
Integration by parts and using Poisson's equation with boundary conditions yields
\begin{equation}
 \delta{\mathfrak F} 
=\int_{\set V}\left(-e\Phi^*+\frac{1}{2}m|\vec{v}|^2
-\bar\epsilon(f^*)\right)\delta f \,d^3v\, d^3r \ .
\end{equation}
Obviously, the necessary condition for a minimum is met by  $f^*(\vec r,\vec v)=\bar{F}\left(\frac{1}{2}m\vec{v}^2 -e\Phi^*(\vec r)\right)$:
The distribution function prescribed at the interface $\set F$ has to hold for the whole domain $\set P$. \linebreak 
Taking into account also Poisson's equation, one arrives exactly at the equilibrium problem.
To show that this equilibrium is indeed stable, consider the second variation and confirm that it is positive definite,
due to the monotonically decreasing function $\bar\epsilon$(f): 
\begin{equation}
 \delta^2{\mathfrak F} 
=\int_{\set V}\left(\frac{1}{2}\epsi|\n\delta\Phi|^2
-\int\frac{1}{2}\bar\epsilon^{\,\prime}(f^*)\,\delta f^2\,d^3v\right)\,d^3r\ge 0 \ .
\end{equation}


\section{Linearized kinetic model}\label{sec:lin}

As we have just seen, the unperturbed equilibrium of the plasma-probe system is stable, 
and it is only the applied electrode voltages that drive the system. We now assume that these \linebreak
voltages are small compared to the thermal voltage $T_{\rm e}/e\approx 3\,{\rm V}$.
(In typical APRS set-up, \linebreak the~applied voltages are much smaller.)
It is then adequate to linearize the dynamic equations around the stationary equilibrium. 
We will assume that the distribution function $f$ in $\set P$ can be described by the equilibrium distribution 
plus a small perturbation  
\begin{align}
	f(\vec r,\vec v,t)	 = 
			\bar{F}\left(\frac{1}{2} m \vec{v}^2 - e\bar{\Phi}(\vec r)\right) + w(\vec r,\vec v) g(\vec r,\vec v,t),\hspace{1cm} | g(\vec r,\vec v,t)| \ll T_{\rm e}.
\end{align}
Here, $w$ is a positive weighting function, defined as the negative derivative of the equilibrium 
distribution function $\bar{F}$
with respect to its argument $\bar\epsilon$:
\begin{align}
		w(\vec r,\vec v) = -\bar{F}^\prime\left(\frac{1}{2} m \vec{v}^2 - e\bar{\Phi}(\vec r)\right).
\end{align}
Also the potential $\Phi(\vec r,t)$ is split up into the equilibrium value $\bar\Phi(\vec r)$ and a perturbation, 
\begin{align}
	\Phi(\vec r,t) = \bar\Phi(\vec r) + \phi(\vec r,t),\hspace{1cm} |\phi(\vec r,t)| \ll T_{\rm e}/e.
\end{align}
The perturbation of the distribution is governed by the linearized kinetic equation  
\begin{equation}
\frac{\p g}{\p t}+ \vec{v}\sca\n g +  \frac{e}{m}\n\bar\Phi\sca\n_{v}g 
-e\vec v\sca\n{\phi}  = - \nu_0  g + \int_\Omega\frac{d\nu}{d\Omega}   g(|\vec v|\vec e)\, d\Omega\ .
\label{BoltzmannLin1}
\end{equation}
Boundary conditions are specular reflection at $\set S$ and (yet undefined) ``smallness''  at $\set F$:
\begin{equation}
  g(\vec r, \vec v, t)
= \left\{\ 
\begin{matrix} 
g(\vec r,\vec v - 2 \vec{n}\!\cdot\!\vec{v} \,\vec{n},t) & & \vec r & \in & \set S\\[1.0ex] 
\ds  g_{\set F}(\vec r, \vec v, t)  & & \vec r & \in & \set F 
\end{matrix}
  \right.,
  \label{BoundaryConditions}
\end{equation}
The perturbation of the potential follows the linearized Poisson equation,
\begin{equation}
 -\nabla\!\cdot\!\left(\epsi(\vec{r})\n\phi\right) 
= \left\{\ 
\begin{matrix} 
0 & & \vec r & \in & \set D\\[1.0ex] 
\ds -e \int_{\R^3} w(\vec r,\vec v) g(\vec r,\vec v,t)\,d^3v  & & \vec r & \in & \set P 
\end{matrix}
  \right.,
  \label{Poisson1}
\end{equation}
together with the boundary conditions (again, ${\phi}_{\set F}$ is small but yet unspecified)
\begin{align}
	&\phi(\vec{r})
= \left\{\ 
\begin{matrix} 
U_n(t) & & \vec r & \in & {\set E}_n,\;\;\; n\in[0,N]\\[0.5ex] 
\ds  {\phi}_{\set F}\left(\vec r,t\right)  & & \vec r & \in & \set F 
\end{matrix}
  \right. \ , \\[0.5ex]
 &  \vec n\!\cdot\!\nabla \phi = 0, \hspace{1.7 cm} \vec r  \in  \set I   \nonumber.
\end{align}

For a more compact notation, we define the  Green's function $G(\vec r, \vec r^\prime)$ as the solution of Poisson's equation for a
unit charge at $\vec{r}^\prime \in \set{V}$ under homogeneous boundary conditions:
\begin{align}
	-&\nabla\sca\left( \varepsilon(\vec{r}) \,\nabla G(\vec r, \vec r^\prime)\right)  = \delta^{(3)}(\vec r -\vec r^\prime), \\
	&G(\vec r, \vec r^\prime) = 0, \hspace{1.5cm}  \vec{r}\in \bigcup_{n=0}^N\,{\set E}_n\cup\set F,\; \vec{r}^\prime \in \set{V}, \nonumber\\[0.5ex]
 &  \vec n\!\cdot\!\nabla G(\vec r, \vec r^\prime) = 0, \hspace{0.7cm} \vec r  \in  \set I   \nonumber.
\end{align}
The Green's function can be shown to be symmetric in its arguments,  
\begin{align}
	  G(\vec r, \vec r^\prime) = G(\vec r^\prime,\vec r)  \hspace{0.7cm} \vec{r}, \vec{r}^\prime  \in  \set V.
\end{align}
We can then formulate the formal solution of the Poisson problem as 
\begin{align}
\phi(\vec r,t) = &
-\int_{\set V} G(\vec r, \vec r^\prime) e \int_{\R^3} w(\vec r^\prime,\vec v) g(\vec r^\prime,\vec v,t)\,d^3v \, d^3 r^\prime \label{FormalPoissonSolution}\\[1.0ex]
&- \sum_{n=1}^N U_n(t) \int_{\set E_n}\varepsilon(\vec r^\prime)\nabla^\prime G(\vec r,\vec r^\prime)\sca d^2{\vec r^\prime}-
\int_{\set F} \phi_{\set F}(\vec r^\prime,t)\,\varepsilon(\vec r^\prime)\nabla^\prime G(\vec r,\vec r^\prime)\sca d^2{\vec r^\prime}.  	\nonumber
\end{align}

The first term of this expression will be called inner potential $\mathsf\Phi$. It is a function of $\vec r$\linebreak  and $t$ and a linear functional of the distribution function $g$:
\begin{align}
	\mathsf\Phi\{g\}(\vec r,t) = -\int_{\set V} G(\vec r, \vec r^\prime)  \int_{\R^3} e w(\vec r^\prime,\vec v)
	g(\vec r^\prime,\vec v,t)\,d^3v \, d^3 r^\prime.
\end{align}
The inner potential obeys Poisson's equation under homogeneous boundary conditions:
\begin{align} 
 -&\nabla\!\cdot\!\left(\epsi(\vec{r})\n\mathsf\Phi\right) 
= \left\{\ 
\begin{matrix} 
0 & & \vec r & \in & \set D\\[1.0ex] 
\ds -e \int_{\R^3} w(\vec r,\vec v) g(\vec r,\vec v,t)\,d^3v  & & \vec r & \in & \set P 
\end{matrix}
  \right.,
  \label{Poisson1}\\
	&\mathsf\Phi(\vec{r}) =0,\hspace{1.8 cm}\vec{r} \in \bigcup_{n=0}^N\,{\set E}_n\cup\set F, \nonumber \\[0.5ex]
 &  \vec n\!\cdot\!\nabla \mathsf\Phi = 0, \hspace{1.6cm} \vec r  \in  \set I   \nonumber.
\end{align}
Another important quantity is the time derivative ${\dot{\mathsf\Phi}}$ of the inner potential; using the kinetic equation it can be
established as a linear functional of $g$, 
\begin{align}
	{\dot{\mathsf\Phi}}\{g\}(\vec r,t) = \int_{\set V} G(\vec r, \vec r^\prime)\, \nabla^\prime \sca \int_{\R^3} e {\vec v} w(\vec r^\prime,\vec v) g(\vec r^\prime,\vec v,t)\,d^3v \, d^3 r^\prime.
\end{align}
It can be used to define the inner current density $\mathsf J\{g\}$, a divergence-free quantity which contains the electron current and the 
displacement current connected to ${{\mathsf\Phi}}$: 
\begin{equation}
 \mathsf J\{g\}(\vec r,t)
=\left\{\ 
 \begin{matrix} 
 \ds -\epsi\n{\dot{\op\Phi}}\left\{g \right\} & & \vec r & \in & \set D\\[1ex] 
 \ds - e\int_{\R^3} \vec v\, g\, w\,d^3v 
     -\epsi\nabla{\dot{\op\Phi}}\left\{g\right\} & & \vec r & \in & \set P 
 \end{matrix}
 \right. \ .
 \label{InnerTotalCurrent}
\end{equation}

The second term of the expression \eqref{FormalPoissonSolution} is the ``vacuum potential''. 
To write it concisely, we define for all $\vec{r}^\prime\in \partial \set{V}$ the characteristic function derived from the Green's function,
\begin{align}
	   \Psi_{{\boldsymbol{r}}^\prime}(\vec r) = - \varepsilon(\vec{r}^\prime)\,\vec{n}(\vec{r}^\prime)\sca\nabla^\prime G(\vec r,\vec r^\prime).
\end{align}
It obeys the Laplace equation under the problem-specific boundary conditions
\begin{align}
 -&\nabla\sca\left(\epsi(\vec r)\nabla \Psi_{{\boldsymbol{r}}^\prime}(\vec r) \right) = 0, \;\;\;\; 
 \vec{r}\in\set{V},  \label{PoissonVacuum}\\
	&\Psi_{{\boldsymbol{r}}^\prime}(\vec r) = \delta^{(2)}(\vec r-\vec{r}^\prime), \;\;\;\;\;\vec{r}\in \bigcup_{n=0}^N\,{\set E}_n\cup\set{F},\nonumber\\
 &  \vec n\!\cdot\!\nabla  \Psi_{{\boldsymbol{r}}^\prime}(\vec r)  = 0, \hspace{1.55cm} \vec r  \in  \set I  \nonumber.
\end{align}
Following \cite{lapke2013}, we also define for all $n\in[0,N]$ the influence functions of the electrodes 
\begin{align}
	\psi_n(\vec r) = \int_{\set E_n} \Psi_{{\boldsymbol{r}}^\prime}(\vec r)\,d^2r^\prime =
		- \int_{\set E_n}\!\varepsilon(\vec r^\prime)\nabla^\prime G(\vec r,\vec r^\prime)\sca d^2{\vec r^\prime}.  
\end{align}
They obey the relations
\begin{align}
 -&\nabla\sca\left(\epsi(\vec r)\n\psi_n\right) = 0,   \label{PoissonVacuum}\\
	&\psi_n  = \left\{\ 
\begin{matrix} 
\delta_{nn^\prime}(\vec r), & & \;\;\;\vec r & \in {\set E}_{n^\prime},  &\;\;\; n^\prime\in[0,N] \\ \nonumber
  0, & &\;\;\;\vec r & \in \set F & 
\end{matrix}
  \right. \ , \\
 &  \vec n\!\cdot\!\nabla \psi_n   = 0, \hspace{1.35cm} \vec r  \in  \set I   \nonumber.
\end{align}
Using these definitions, the vacuum potential can be written as
\begin{align}
\phi^{({\rm vac})}(\vec r,t)  =  \sum_{n=1}^N U_n(t) \psi_n(\vec r)+ 
\int_{\set F} \phi_{\set F}(\vec r^\prime,t) \Psi_{{\boldsymbol{r}}^\prime}(\vec r)\, d^2 r^\prime.
\end{align}
The displacement current related to the vacuum potential is
\begin{align}
	\vec{J}^{({\rm vac})}(\vec r, t) =  -\sum_{n=1}^{N}\frac{\p U_n}{\p t}\epsi(\vec r)\n\psi_n 
	-  \!\int_{\set F}\frac{\partial\phi_{\set F}}{\partial t}(\vec r^\prime)
	 \epsi(\vec r)\nabla\Psi_{{\boldsymbol{r}}^\prime}(\vec r) \, d^2{r}^\prime.
\end{align}

Of particular importance are the currents through the electrodes and the outer boundary. 
For the inner contributions at the electrode ${\set E}_n$, we obtain (where in the second step  we employ an argument
outlined in the appendix of \cite{lapke2013}):
\begin{align}
	 {\mathsf I}_n\{g\}(t) = - \int_{{\set E}_n}\mathsf J\{g\}  \sca d^2\vec r =  \int_{\set P}\int_{\R^3} e\nabla\psi_n \sca  \vec v\, g\,w\,d^3v\,d^3r, \hspace{0.55cm} n \in [0,N],
\end{align}
through the outer boundary, the current is
\begin{align}
{\mathsf I}_{\set F}\{g\}(t)  = - \int_{{\set F}}\mathsf J\{g\}  \sca d^2\vec r =   \int_{{\set F}}  \left( e\int_{\R^3} \vec v\, g\, w\,d^3v  
     +\epsi\nabla{\dot{\op\Phi}}\left\{g\right\} \right) \sca d^2\vec r = -\sum_{n=0}^N  {\mathsf I}_n\{g\}(t).
\end{align}

\pagebreak

To write the vacuum currents concisely, we define the electrode capacitance matrix $C_{nn'}$, 
the electrode boundary coupling $c_n(\vec r)$,  and the boundary-boundary self-coupling $c(\vec{r},\vec{r}^\prime)$, 
\begin{align}
	C_{nn'} &= \int_{\set V} \epsi(\vec r) \nabla\psi_n \sca \nabla\psi_{n^\prime} \, d^3 r \equiv 
	    \int_{\set{E}_{n^\prime}} \epsi(\vec r) \nabla\psi_n \sca d^2\vec{r}
		, \hspace{2.85cm} n,n^\prime \in [0,N],\\[0.0ex]
	c_n(\vec r)&=  \int_{\set V}\epsi(\vec r^\prime )\nabla^\prime \psi_{n}\sca 
	                 \nabla^\prime  \Psi_{{\boldsymbol{r}}}(\vec{r}^{\prime})\, d^3 r^\prime \equiv \varepsilon(\vec{r})\, \vec{n} \sca \nabla \psi_n(\vec{r}),
									\hspace{2.05cm} n \in [0,N], \vec{r} \in \set{F},\\[0.0ex]
	c(\vec{r},\vec{r}^\prime) &= 	\int_{\set V} \varepsilon(\vec r^{\prime\prime})\nabla^{\prime\prime}
	\Psi_{{\boldsymbol{r}}}(\vec{r}^{\prime\prime})\sca \nabla^{\prime\prime}  \Psi_{{\boldsymbol{r}}^\prime}(\vec{r}^{\prime\prime})\, d^3 r^{\prime\prime} 
	\equiv \varepsilon(\vec{r})\, \vec{n} \sca \nabla   \Psi_{{\boldsymbol{r}}^\prime}(\vec r),
	\hspace{0.85cm} \vec{r},\vec{r}^\prime \in \set{F}.
\end{align}
They obey the identities
\begin{align}
&\sum_{n^\prime=0}^N C_{nn'} + \int_{\set F} c_n(\vec r)\, d^2 r = 0, \hspace{0.85cm} n \in [0,N], \\
&\sum_{n=0}^N c_n(\vec r) + \int_{\set F} c(\vec r,\vec{r}^\prime)\, d^2 r = 0,\hspace{0.55cm} \vec{r}^\prime \in\set F\ .
\end{align}
The vacuum currents through the electrodes ${\set E}_n, n \in [0,N]$, are then
\begin{align}
	I^{({\rm vac})}_n(t) =- \int_{{\set E}_n}\vec{J}^{({\rm vac})} \sca d^2\vec r  =  \sum_{n'=1}^N C_{nn'}\frac{dU_{n'}}{dt} + \int_{\set F}c_n(\vec r) \frac{\partial\phi_{\set F}}{\partial t}(\vec r) \, d^2 r,
\end{align}
the vacuum current density and the total vacuum current through the interface $\set F$ are
\begin{align}
&J^{({\rm vac})}(\vec r, t)  = \vec{n}\sca	\vec{J}^{({\rm vac})}(\vec r, t) 
	=  -\sum_{n=1}^{N}\frac{\p U_n}{\p t} c_n(\vec{r})
	-  \int_{\set F} c(\vec{r},\vec{r}^\prime)\frac{\partial\phi_{\set F}}{\partial t}(\vec r^\prime) d^2 r^\prime,\\
&I_{\set F}^{({\rm vac})}(t) =  \int_{{\set F}}\vec{J}^{({\rm vac})}(\vec r, t)   \sca d^2\vec r	=
 \int_{{\set F}}J^{({\rm vac})}(\vec r, t)   d^2 r = \sum_{n=0}^N I^{({\rm vac})}_n(t).
\end{align}

We now turn to the calculation of the electric field energy. The contributions from the inner field and the
vacuum field decouple, as can be shown by partial integration,
\begin{align}
	\int_{\set V} \frac{1}{2}\varepsilon(\vec r)\nabla\phi^2\, d^3 r = 
	\int_{\set V} \frac{1}{2}\varepsilon(\vec r)\nabla\mathsf\Phi\{g\}^2\, d^3 r +
	\int_{\set V} \frac{1}{2}\varepsilon(\vec r)\nabla{\phi^{({\rm vac})}}^2\, d^3 r. 
\end{align}
In terms of the capacitive coupling coefficients, the vacuum energy can be expressed as
\begin{align}
	 \int_{\set V} \frac{1}{2}\varepsilon(\vec r)\nabla{\phi^{({\rm vac})}}^2\, d^3 r = &
	\frac{1}{2} \sum_{n=1}^N \sum_{n^\prime=1}^N  C_{nn^\prime} U_n U_{n^\prime} +
\sum_{n=1}^N U_n(t)  
\int_{\set F}  c_{n}(\vec r)\phi_{\set F}(\vec r)d^2{r}\,\\
&+\frac{1}{2} \int_{\set F}\int_{\set F} c(\vec{r},\vec{r}^\prime) \phi_{\set F}(\vec r,t)  \phi_{\set F}(\vec r^\prime,t) \,d^2r\,d^2{r^\prime}\nonumber.
\end{align}
For the time derivatives of the inner field energy and the vacuum energy, we obtain the following relations
which also demonstrate their decoupling:
\begin{align}
&\frac{d}{dt}\int_{\set V} \frac{1}{2}\varepsilon(\vec{r}) \n{\mathsf\Phi}\left\{g\right\}^2\, d^3 r = -\int_{\set P} \n{\mathsf\Phi}\left\{g\right\}\sca e\int_{\R^3} \vec v\, g\, w\,d^3v \, d^3r,\\[0.5ex]
&\frac{d}{dt}\int_{\set V} \frac{1}{2}\varepsilon(\vec r)\nabla{\phi^{({\rm vac})}}^2\, d^3 r = 
 \sum_{n=1}^N U_n I^{(\rm vac)}_n - \int_{\set F} \phi_{\set F}(\vec r)\,J^{(\rm vac)}(\vec r) \,d^2r  \nonumber.
\end{align}

As a last point in this section, we now consider the balance of the linearized free energy.
Of course, the analysis of section \ref{sec:thermo} applies, but it is more instructive to rederive the results employing the 
linearized kinetic equation of this section. With all definitions substituted, and the vacuum field 
written on the right, it reads:    
\begin{align}
\frac{\p g}{\p t}+ \vec{v}\sca\n g + & \frac{e}{m}\n\bar\Phi\sca\n_{v}g 
-e\vec v\sca\n{\mathsf\Phi}\left\{g\right\} \\& = - \nu_0  g
+\int_\Omega\frac{d\nu}{d\Omega}   g(|\vec v|\vec e)\, d\Omega  
+\sum_{n=1}^N   U_n e\vec v\sca\n\psi_{\rm n} +e\vec{v}\sca \int_{\set F} \phi_{\set F}(\vec r^\prime,t) \nabla\Psi_{{\boldsymbol{r}}^\prime}(\vec r)\, d^2 r^\prime. \nonumber
\label{LinearizedBoltzmannEquation}
\end{align}
We multiply this by $g\,w$ and integrate over the velocity space $\R^3$ and the plasma domain $\set P$.\linebreak  Taking into account that the collision term integral has a definite sign,
as shown by \eqref{HTheorem}, \linebreak the current relation \eqref{InnerTotalCurrent},
the fact that $w$ vanishes for $|\vec v|\to\infty$, and the specular reflection\linebreak  boundary conditions for $g$ at $\set S$, we obtain
\begin{align}
\frac{d}{dt} \int_{\set P}\int_{\R^3} \frac{1}{2} g^2 \, w \, d^3v\, d^3r 
&+ \int_{\set F}\int_{\R^3}  \vec{v} \frac{1}{2} g_{\set{F}}^2 w  \, d^3v\, \sca\, d^2\vec{r}
- \int_{\set P}\n{\mathsf\Phi}\left\{g\right\}\sca \int_{\R^3} g\,e\vec v
 \, w \, d^3v\, d^3r  \\
&\le  \sum_{n=1}^N   U_n {\mathsf I}_n\{g\} - \int_{\set F} {\phi}_{\set F}\left(\vec r\right)\mathsf J\{g\}(\vec r,t) \!\cdot\! d^2\vec r. \nonumber
\end{align}
Adding the inner potential energy balance and re-arranging terms yields
\begin{align}
 \frac{d}{d t}&\left(\int_{\set V} \frac{1}{2}\varepsilon(\vec r)\nabla\mathsf\Phi\{g\}^2\, d^3 r  + 
 \int_{\set P}\int_{\R^3}     \frac{1}{2}w g^2\, d^3v\,d^3r \right)\\[1ex]
& + \int_{\set F}\left( \delta {\Phi}_{\set F}\left(\vec r\right)\mathsf J\{g\}(\vec r,t) 
  +\int_{\R^3} {\vec v} \frac{1}{2} w g_{\set{F}}^2  d^3v\right)\!\cdot\! d^2\vec r 
\le \sum_{n=1}^N U_n  {\mathsf I}_n\{g\}(t).\nonumber
\end{align}
The second term on the left represents the excess entropy exchanged with the environment through the interface $\set F$.
As stated above, it is our \textit{physical postulate} that this quantity has a definite sign; i.e., 
cannot become negative:
\begin{align}
	\int_{\set F}  \Delta\vec\Gamma_{\sigma\set{F}}  \sca d^2\vec{r} = \int_{\set F}\left( {\phi}_{\set F}\left(\vec r\right)\mathsf J\{g\}(\vec r,t) 
  +\int_{\R^3} {\vec v} \frac{1}{2} w g_{\set F}^2  d^3v\right)\!\cdot\! d^2\vec r \ge 0.
\end{align}

The kinetic free energy $\mathbb{F}\{g\}$ suggested by these relations is a quadratic, positive definite 
functional of the  distribution perturbation $g$. It is identical to the second variation of $\mathfrak{F}$, except that the contribution of the vacuum energy is not included:
\begin{align}
	  \mathbb{F}\{g\} :=  \int_{\set P}\int_{\R^3} \frac{1}{2} g^2 \, w \, d^3v\, d^3r
+\int_{\set V}\frac{1}{2}\varepsilon(\vec{r})\nabla{\mathsf\Phi}\left\{g\right\}^2  \,d^3r.
\end{align}
For this quantity, we have established one of the main results of this manuscript,
\begin{align}
 \frac{d  }{dt}   \mathbb{F}\{g\} 
\le \sum_{n=1}^N U_n  {\mathsf I}_n\{g\}(t).
\end{align}

\pagebreak


\section{Functional analytic description}\label{sec:func}

We will now continue our analysis of APRS by transforming the ``physical'' description of the plasma-probe system 
into a ``mathematical'' model. To achieve this, we must give exact meaning to the physical assumptions made above.
These were the length scale ordering 
$\lambda_{\rm D} \lu \lambda \approx R \ll 	R_{\set V} \ll L \approx \lambda_{\epsilon} \approx \lambda_{\rm s}$, the 
time scale ordering $\omega_{\rm pe} \gu \omega_{\rm RF}\gu \nu_0 \gg \nu_{\rm i} \approx  \omega_{\rm pi} \gg \omega_{\rm g}$, \linebreak  
and the boundary conditions and physical postulates at the influence domain interface $\set F$. 
In detail, we proceed as follows:
\begin{enumerate}
	\item The scales $L$ (reactor dimension), $\lambda_{\epsilon}$ (energy diffusion length), and $\lambda_{\rm s}$ (skin depth)
	      are not considered any longer finite but infinite. Similarly, the frequencies $\nu_{\rm i}$ (inelastic collision frequency), $\omega_{\rm pi}$
				(ion plasma frequency), and $\omega_{\rm g}$ (neutral dynamics frequency) are set equal to zero.
				There are no consequences for our model; these scales and the corresponding processes 
				(gradual establishment of the nondissipative equilibrium) were not considered at all. 
				(These scales are simply not ''observed'' by the probe.)
	
	\item The scale $R_{\set V}$ is set to infinity: We enlarge the finite influence domain $\set V$ to be infinite. \linebreak
	      Of course, the ``thermodynamic'' arguments of section \ref{sec:thermo} will then no longer apply; \linebreak they rely on the
				assumption that the influence domain is in contact with the unperturbed plasma environment via the interface $\set F$.  
				The validity of the linear model, however, is not affected; the equilibrium distribution $\bar F(\epsilon)$ 
				is already incorporated. Technically, the interface $\set F$ moves to infinity, 
				the boundary values $g_{\set F}$  and $\phi_{\set F}$ vanish, and the functions $c_n(\vec{r})$ and 
				$c(\vec{r},\vec{r}^\prime)$ loose their meaning.  Physically, ${\set F}$ assumes the role of a distant ground and is treated as such.
				We can assume that the equilibrium plasma at large distances is homogeneous.
				
	\item There is no formal ordering assumed of the remaining spatial scales $\lambda_{\rm D}$ (Debye length), 
	      $\lambda$ (elastic electron mean free path), 
	      and $R$ (probe scale), nor of the frequency scales\linebreak $\omega_{\rm pe}$ (plasma frequency), $\omega_{\rm RF}$ (applied RF frequency),
				and $\nu_0$ (collision frequency). \linebreak
				All these scales will be taken as finite; the statements $\lambda_{\rm D} \lu \lambda \approx R $ and 
				$\omega_{\rm pe} \gu \omega_{\rm RF}\gu \nu_0$ merely indicate the ``typical'' APRS situation. 
				The collision free limit $\lambda\to\infty$, $\nu_0\to 0$, where kinetic effects dominate over collisional effects (see introduction)  
				will not be excluded from the description.
\end{enumerate}
 
We now summarize our model of the probe-plasma system. Its core is a linear kinetic equation with appropriate boundary conditions for the 
distribution perturbation $g$:
\begin{align} 
&\frac{\p g}{\p t}+ \vec{v}\sca\n g +  \frac{e}{m}\n\bar\Phi\sca\n_{v}g 
-e\vec{v}\sca\n{\mathsf\Phi}\left\{g\right\} =- \nu_0  g
+\int_\Omega\frac{d\nu}{d\Omega}   g(|\vec v|\vec e)\, d\Omega  =\sum_{n=1}^N   U_n e\vec v\sca\n\psi_{\rm n}, 
\label{LinearizedBoltzmannEquationInfinite} \\
&\;\;\;\;\;  g(\vec r, \vec v, t)
= \left\{\ 
\begin{matrix} 
g(\vec r,\vec v - 2 \vec{n}\!\cdot\!\vec{v} \,\vec{n},t) & & \vec r & \in & \set S\\[0.0ex] 
\ds  0  & & \vec r & \to & \infty \label{SpecBC}
\end{matrix}
  \right..\end{align}
The inner potential $\mathsf\Phi$ is a function of $\vec r$ and $t$ and a homogeneous, linear functional of $g$:
\begin{align}
	\mathsf\Phi\{g\}(\vec r,t) = -\int_{\set V} G(\vec r, \vec r^\prime)  \int_{\R^3} e 
	g(\vec r^\prime,\vec v,t)\,w(\vec r^\prime,\vec v)\,d^3v \, d^3 r^\prime,
\end{align}
where the Green's function $G(\vec r, \vec r^\prime)$ obeys:
\begin{align}
	-&\nabla\sca\left( \varepsilon(\vec{r}) \,\nabla G(\vec r, \vec r^\prime)\right)  = \delta^{(3)}(\vec r -\vec r^\prime), \\
	&G(\vec r, \vec r^\prime) = 0, \hspace{1.5cm}  \vec{r}\in \bigcup_{n=0}^N\,{\set E}_n\;\;{\rm or}\;\; \vec{r}\to \infty, \;\; \vec{r}^\prime \in \set{V}, \nonumber\\[0.5ex]
 &  \vec n\!\cdot\!\nabla G(\vec r, \vec r^\prime) = 0, \hspace{0.7cm} \vec r  \in  \set I   \nonumber.
\end{align}
The RF excitation is represented by the electrode functions $\psi_n, n\in [1,N]$,
\begin{align}
 -&\nabla\sca\left(\epsi(\vec r)\n\psi_n\right) = 0,    \label{PoissonVacuum}\\
	&\psi_n  = \left\{\ 
\begin{matrix} 
\delta_{nn^\prime}(\vec r), & & \;\;\;\vec r & \in & {\set E}_{n},\\ \nonumber
  0, & &\;\;\;\vec r & \to & \infty 
\end{matrix}
  \right. \ , \\
 &  \vec n\!\cdot\!\nabla \psi_n   = 0, \hspace{1.35cm} \vec r  \in  \set I   \nonumber.
\end{align}
The inner currents and the vacuum currents  through the electrodes are, for $n \in [1,N]$,
\begin{align}
	 &{\mathsf I}_n\{g\}(t) =  \int_{\set P}\int_{\R^3} e\nabla\psi_n \sca  \vec v\, g\,w\,d^3v\,d^3r\\
   &I^{({\rm vac})}_n(t) =  \sum_{n'=1}^N C_{nn'}\frac{dU_{n'}}{dt},
\end{align}
with the capacitive coefficients calculated as follows, for $\ n,n^\prime \in [1,N]$: 
\begin{align}
		C_{nn'} = \int_{\set V} \epsi(\vec r) \nabla\psi_n \sca \nabla\psi_{n^\prime} \, d^3 r \equiv 
	    \int_{\set{E}_{n^\prime}} \epsi(\vec r) \nabla\psi_n \sca d^2\vec{r}.
\end{align}
A quadratic free energy functional was established with a definite time derivative:
\begin{align}
	  \mathbb{F}\{g\} :=  &\int_{\set P}\int_{\R^3} \frac{1}{2} g^2 \, w \, d^3v\, d^3r
+\int_{\set V}\frac{1}{2}\varepsilon(\vec{r})\nabla{\mathsf\Phi}\left\{g\right\}^2  \,d^3r,\\[0.5ex]
 &\frac{d  }{dt}   \mathbb{F}\{g\} 
\le \sum_{n=1}^N U_n  {\mathsf I}_n\{g\}(t).
\end{align}
  
To get deeper insight into the model, we now proceed to establish an abstract description. \linebreak
The appropriate framework for this is functional analysis.
The set of all possible distribution functions $g(\vec r,\vec v)$ on the phase space $\set P\times\mathbb{R}^3$ naturally forms 
a linear configuration space. \linebreak To turn it into a Hilbert space $\set H$, a scalar product is required and the completition process must be carried out.
The following choice will result in a weighted $L^2$-space: 
\begin{align}
	(g^{\prime*}| g ) :=  \int_{\set P}\int_{\R^3} g^{\prime*} g\, w \, d^3v \,d^3r\ \label{SP}. 
\end{align}
For our purposes, however, it is more suited to employ a scalar product motivated by the linearized free energy.
The following definition meets all aspects of an inner product, \linebreak namely i) conjugate symmetry, ii) sesquilinearity, and iii) positive definiteness: 
\begin{align}
	\! \langle g^\prime | g \rangle := \int_{\set P}\int_{\R^3}  g^{\prime*} g\, w \, d^3v\, d^3r
+\int_{\set V}\varepsilon(\vec{r})\nabla{\mathsf\Phi}\{g^{\prime*}\}\sca
\nabla{\mathsf\Phi}\left\{ {g}\right\}  \,d^3r. 
\label{ScalarProduct}
\end{align}
Integrating the second term by parts and utilizing Poisson's equation one finds
\begin{align}
\langle g'|g\rangle 
= \int_{\set P}\int_{\R^3} g^{\prime*}  g\, w \, d^3v \,d^3r 
 - \int_{\set P}\op\Phi\{g^{\prime*}\}\int_{\R^3}ewg\,d^3v\,d^3r 
= (g^\prime | g )  \; 
 + ( \textstyle {-e\mathsf\Phi}\left\{g'\right\}| g ) \ .
\label{ScalarProduct2} 
\end{align}
Both scalar products are related; $\langle . | . \rangle$ may be called the energetic scalar product 
associated to the ``Coulomb integral'' operator $-\mathsf{\Phi}$. 
In the usual way, the inner product  induces a norm; its square corresponds to the quadratic free energy 
up to a factor of two:
\begin{align}
	||g||^2 := \langle g | g \rangle = 2\, \mathbb{F} \{g\}.
\end{align}

Within the state space  $\set H$, the dynamics can be formulated as a differential equation for the 
dynamic state vector $g$. We introduce the excitation state vectors $e_n =e\vec v\sca\n\psi_n$
and two dynamic operators, the Vlasov operator $\op T_V$ and the collision operator $\op T_S$:  
\begin{eqnarray}
\op T_V g 
& = &- {\vec v}\sca\n g - \frac{e}{m}\n\bar\Phi\sca\n_v g
      +e\n\op\Phi\left\{g\right\}\sca \vec v,   \\[1ex]
\op T_S g 
& = & -\nu_0 g+\int_{\Omega} \frac{d\nu}{d\Omega}\, g(|\vec v|\vec{e})\,d\Omega.
\end{eqnarray}
The dynamical equation then assumes the form
\begin{equation}
\frac{\p g}{\p t}=\op T_V g +\op T_S g + \sum_{n=1}^N U_n e_n.
\label{DynKin1}
\end{equation}
The response of the system, the inner current, is as follows; the excitations vectors hence also serve as observation vectors:
\begin{align}
	{\mathsf I}_n\{g\}(t) =  \int_{\set P}\int_{\R^3} e\vec v\sca\nabla\psi_n   \, g\,w\,d^3v\,d^3r = 
	( e_n | g ) = \langle e_n | g \rangle.
\end{align}

The behavior of the system depends crucially on the properties of the dynamic operators.
A detailed analysis will be presented in the appendix; here only a short summary is given. \linebreak
The operator $\op T_V$ contains derivatives with respect to $\vec{r}$ and $\vec{v}$ and is therefore \textit{unbounded},
i.e., there is a family of states $g_\gamma$ in $\set H$ whose norm $||g_\gamma||$ is unity  but whose images under the \linebreak operator diverge,
$\lim_{\gamma\to\infty} ||\op T_V g_\gamma||=\infty$.
The domain $\set D(\op T_V)$ is thus only a dense subset of $\set H$,\linebreak namely the set of distribution functions $g$ which are differentiable 
and obey the specular boundary conditions at the surface $\set{S}$. The adjoined operator ${\op T}_V^*$ has the same 
 domain $\set D({\op T}_V^*)=\set D(\op T_V)$ and  
	 is identical to $-\op T_V$. This property is called \textit{skew self-adjointness} and implies that for two distribution functions 
	$g$ and $g^\prime$ in $\set D(\op T_V)$ we have 
	 \begin{align}
		  \langle\op T_V\, g^\prime|g\rangle= -\langle g^\prime|\op T_V\, g\rangle.	
		\end{align}
Together, these properties imply that the spectrum of the operator $\op T_V$ is purely imaginary, and reaches from $-i\,\infty$ to $i\,\infty$.
As argued in the appendix, the spectrum is \textit{continuous}.\linebreak 
Then the spectral representation of $\op T_V$ reads as follows,
where $\op P_V(\omega)$ is a resolution of the identity, i.e., a family of projection operators with $\op P_V(-\infty)=0$ and $\op P_V(\infty)=1$:
\begin{align}
	\op T_V = \int_{-\infty}^\infty i \omega \, d\op P_V(\omega) \label{SpectralRepresentationTV}
\end{align}
The collision operator $\op T_S$ is the difference of an integral operator
(with kernel $\frac{d\nu}{d\Omega}(\vec{r},v,\vartheta)$) and a multiplication operator (with $\nu_0(\vec{r},v)$). 
These are regular functions; the operator is thus bounded  
and its domain $\set D(\op T_S)$ is the full 
Hilbert space $\set H$. 
It is further \textsl{symmetric} and therefore \textit{self-adjoint}; for two distributions $g$ and $g^\prime$ in $\set H$ we have
	 \begin{align}
		  \langle\op T_S\, g^\prime|g\rangle=\langle g^\prime|\op T_S\, g\rangle.	
		\end{align}
The kernel $\set K(\op T_S)$ of the operator $\op T_S$ is the set of all distribution functions that are isotropic with respect to the velocity.
Except for those distribution functions, the operator is negative; altogether it is \textit{negative semi-definite}, i.e., for all $g$ in $\set H$
 \begin{align}
		  \langle g|\op T_S\, g\rangle \le 0;	
		\end{align}
Together, these properties imply that the spectrum of the operator $\op T_S$ is real and can be included in the interval $[-\nu_{\rm max},0]$.
The spectral representation of $\op T_S$ thus reads as follows,
where $\op P_S(\nu)$ is another resolution of the identity with $\op P_S(0)=0$ and $\op P_S(\nu_{\rm max})=1$:
\begin{align}
	\op T_S = -\int_{0}^{\nu_{\rm max}} \nu \, d\op P_S(\nu). \label{SpectralRepresentationTS}
\end{align}
 
The characterization of the full operator $\op T = \op T_V+\op T_S$ is the subject of ongoing research.
We assume, however, that its spectrum is entirely located in the negative half plane of $\mathbb{C}$, so~that a   
harmonic ansatz with frequency $\omega_{\rm RF}$ for the excitation and the response is allowed. 
We find the corresponding solution of the equation as
\begin{equation}
g=\sum_{n=1}^N U_n \left(i\omega_{\rm RF}-\op T_V-\op T_S\right)^{-1} e_n.
\label{AlgKinZu}
\end{equation}
The inner current -- the response of the system -- is then  
\begin{equation}
{\mathsf I}_n\{g\} =\sum_{n'=1}^N \bra{e_n}\left(i\omega_{\rm RF}-\op T_V-\op T_S\right)^{-1} e_{n'}\rangle\, U_{n'}=\sum_{n'=1}^N Y_{nn'}U_{n'}
\label{FinalCurrent}.
\end{equation}
Thus, the system response function $Y_{nn'}(\omega_{\rm RF})$ is given in terms of the matrix elements of the resolvent of the dynamic operator evaluated for values on the imaginary axis 
\begin{equation}
Y_{nn'}=\bra{e_n}\left(i\omega_{\rm RF}-\op T_V-\op T_S\right)^{-1} e_{n'}\rangle.
\label{response}
\end{equation}

This is the result promised above: Also in a kinetic model, the spectral response of the probe-plasma system can 
be expressed in terms of  matrix  elements of the resolvent of the dynamical operator.

\pagebreak


\section{Summary and conclusion}\label{sec:conclusion}

In this manuscript we derived and discussed a fully kinetic model of electrostatic APRS (active plasma resonance spectroscopy). 
The subject of our analysis was the interaction of
 an arbitrarily shaped, dielectrically covered  RF probe with the 
plasma of its influence domain $\set V$. \linebreak
On the length scale of the influence domain $R_{\set V}$, and the time scale of the interaction $\omega_{\rm pe}^{-1}$, that plasma was  
assumed to be in a stable, nondissipative equilibrium, characterized by a distribution $\bar f$ which 
is a sole function of the total energy, ${\bar f}(\vec r,\vec v) \equiv \bar{F}\left(\frac{1}{2}m\vec v^{2}-e\bar{\Phi}(\vec r)\right). $\linebreak
(We stressed that nonequilibrium processes on larger length or time scales are not precluded.)
Exploiting a formal   analogy to classical thermodynamics, 
 we defined the  kinetic free energy ${\mathfrak F}$\linebreak of the domain $\set V$ 
as the difference of the total energy $\mathfrak{U}$ and the kinetic entropy $\mathfrak{S}$
 and showed that it has
a definite time derivative with respect to the RF power.

We then turned to a linearized model of the plasma-probe interaction, valid for applied RF voltages smaller 
than $T_{\rm e}/e \approx 3\,{\rm V}$. (A condition well met in standard APRS configurations.)
The fact that the second variation of the free energy is a positive definite functional of the perturbation $g$ 
motivated a
scalar product $\langle g^\prime| g\rangle$ and allowed to define a Hilbert space $\set H$.
The behavior of the plasma-probe system could then be captured by a dynamical equation,
and the response to a harmonic excitation could be expressed by matrix elements of the resolvent of the dynamical operator: 
Equation \eqref{response} is our main result.

Formaly, the derived response function is identical to the corresponding expression of \cite{lapke2013}
 which was obtained on the basis of the cold plasma model. Of course, this raises the question: \linebreak How do the two results compare? In particular, 
what reflects that \eqref{response} is not limited to the regime of relatively high pressure, but holds for
all pressures, including the limit $p\to 0$? 
In a nutshell: The fact that the operator $\op T_V$ here (presumely) has  a continuous spectrum, while the spectrum of
the corresponding operator in ref.~\cite{lapke2013} is discrete.

\begin{figure}[h!]
\hspace{5mm}\includegraphics[width= 0.8\columnwidth]{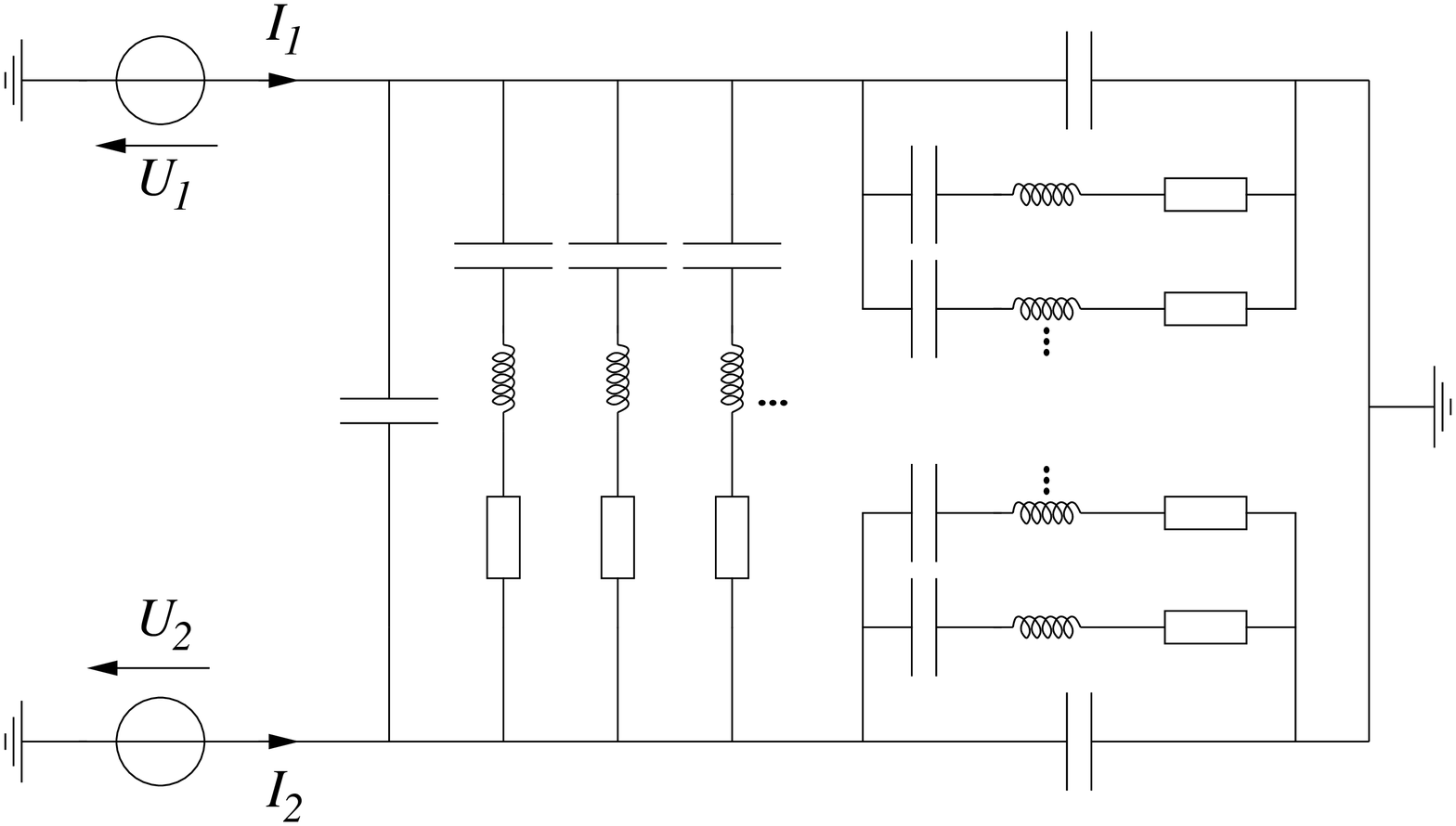}

\includegraphics[width= 0.8\columnwidth]{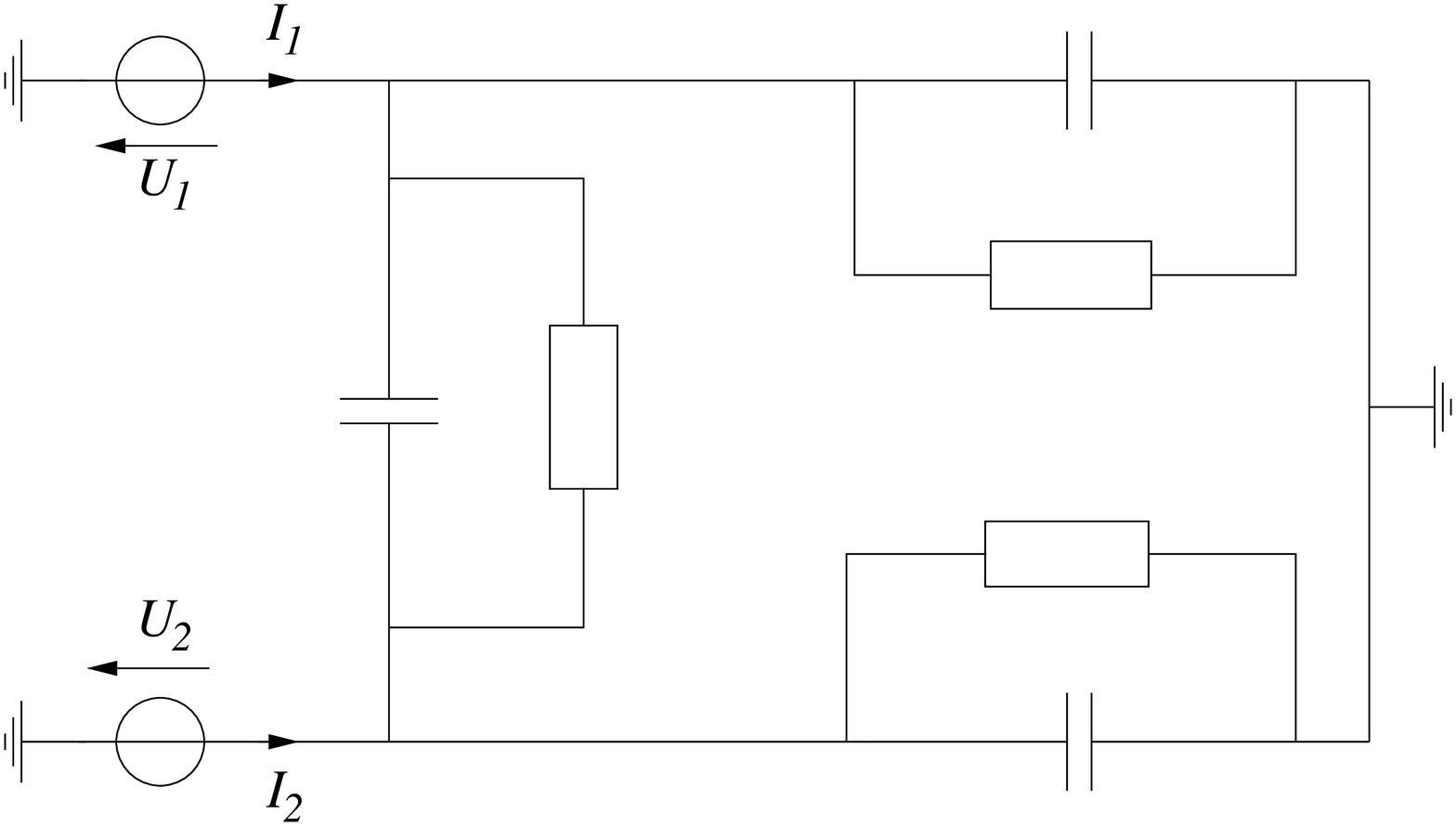}
\caption{Equivalent circuits of electrostatic APRS in the case of a probe with two electrodes,
         illustrating  the differences between the cold plasma model (top, reprint \cite{lapke2013}) and the kinetic model (bottom).
				 Both circuits exhibit the direct coupling branch between the electrodes and two additional branches 
				 which couple to ground. Each branch consists of the vacuum capacitance and the inner coupling. 
				 In the cold plasma model, the inner coupling can be represented by an infinite number of discrete resonance circuits with collisional damping.
				 In the kinetic model, such a separation is not possible; the inner coupling is represented by a complex impedance whose real part
				 reflects both collisional and noncollisional dissipation.}\label{EquivalentCircuit}
\end{figure}

To illustrate the situation, consider the electrically symmetric multipole resonance probe. For this realization of APRS, ref.~\cite{lapke2013} derived the equivalent circuit depicted in fig.~\ref{EquivalentCircuit} (top). 
The circuit has three nodes, namely the two driven electrodes ${\set E}_1$ and ${\set E}_2$, and ground ${\set E}_0$. They are coupled by vacuum capacitances and infinitely many discrete resonance circuits, 
each of which represents an eigenvalue pair of the dynamical operator. The spectral response is thus a rational function,
i.e., a sum of Lorentz  curves. The damping is caused by electron-neutral collisions and vanishes in the limit $p \to 0$, 
the resonance peaks then diverge.

In our kinetic model, the situation is entirely different, as illustrated by fig.~\ref{EquivalentCircuit} (bottom). 
The dynamical operator has no discrete eigenvalues, instead it has a continous spectrum. 
This has important consequences: The inner coupling cannot be decomposed in a sum of discrete resonance circuits 
but must be represented by an integral. The spectral response becomes
a non-rational function of $\omega_{\rm RF}$. In the limit $p \to 0$, 
there are no divergences anymore.\linebreak  Instead, a new phenomenon appears, related to anomalous or non-collisional dissipation. \linebreak 
To see this explicitly, consider the
matrix elements of the response function in the limit of zero electron-neutral collisions. 
Utilizing the spectral representation of the resolvent of $\op T_V$, and introducing a proper regularization,
we can evaluate them as follows:
\begin{align}
	Y_{nn'} &= \lim_{\nu\to 0}\bra{e_n}\left(i\omega_{\rm RF}-\op T_V+\nu\right)^{-1} e_{n'}\rangle\label{response0}\\[1.5ex]
	        &= \lim_{\nu\to 0} \int_{-\infty}^\infty \frac{1}{i\omega_{\rm RF}-i\omega + \nu} \, d\,\langle {e_n} |\, \op P_V(\omega)   e_{n'}\rangle\ .
\nonumber \\[1.5ex]
         &= \frac{1}{i} \dashint_{-\infty}^\infty \frac{1}{\omega_{\rm RF}-\omega} \, d\,\langle {e_n} |\, \op P_V(\omega)   e_{n'}\rangle\  + 
      \pi    \langle {e_n} |\, \op P^\prime_V(\omega_{\rm RF})   e_{n'}\rangle\  = i X_{nn^\prime} + R_{nn^\prime}.
			\nonumber
\end{align}
The last form was obtained by the Plemelj formula. The principal value term is imaginary; \linebreak
the residuum is real, positive definite (as matrix), and describes the anomalous dissipation. 
There is an obvious physical analogy to the radiation damping of an electromagnetic antenna: In a periodic state, the probe constantly emits plasma waves  
which propagate to ``infinity''. 
(These waves will eventually be Landau damped \cite{villani2011}, but the free energy is conserved and will continue to propagate.)
The corresponding distribution $\tilde{g}$ can in principle be calculated but is not square integrable and thus not an element of $\set H$.
However, we may assume that the projection $\langle e_n|\tilde{g}\rangle$ on the observation vectors exists: The free energy simply leaves the 
``observation range'' of the probe.

In summary: We have presented a kinetic functional analytic description of electrostatic 
active plasma resonance spectroscopy including a closed expression for the spectral response. \linebreak
Among other insights, we found an explanation for the experimentally observed collisionless broadening of the spectrum at low pressure. 
Future work will include applying the formalism \linebreak  to concrete APRS probe designs, especially to our own multipole resonance probe (MRP).  \linebreak
Particular emphasis will be placed on comparing eqs.~\eqref{response} and \eqref{response0} with experimental data.
Our ultimate goal is to establish explicite ``formulas'' which will allow to
derive not only the electron density but also the electron temperature and the effective electron collision frequency 
from the measured spectrum.

\pagebreak

\appendix

\section{Properties of the Vlasov operator}\label{sec:AppTV}

We first focus on the Vlasov operator $\op T_V$ which is a differential operator with derivatives both with respect to 
$\vec{r}$ and to $\vec{v}$. As such, it is unbounded. This can be verified by defining a family of test vectors which are bounded but whose images under the operator diverge
\begin{align}
		g_\gamma =  g_\gamma(\vec r,\vec v) = \sin\left(\gamma \frac{\vec{n}\sca \vec{v}}{\hat v}\right)\,\frac{1}{{\left(\pi \hat{v}^2\right)}^{3/4}}\,
		\exp{\left(-\frac{\vec{v}^2}{2 \hat v^2} \right)} \sqrt{\frac{h(\vec r)}{w(\vec r,\vec v)}}.
\end{align}
Here, $\vec{n}$ is a given unit vector, $\hat v$ a velocity scale, and $h(\vec{r})$ a smooth non-negative function with support in $\set V$,
normalized to $	\int_{\set P} h(\vec r)\,d^3r = 1$.
The inner potential $\op \Phi\{g_\gamma\}$ equals zero, \linebreak as the function $g_\gamma$ is odd in $\vec{v}$ and has no charge density.
Thus, the norm of the test state can be computed as follows; it is bounded for arbitrary $\gamma>0$:
\begin{align}
 \norm{g_\gamma}^2
=\int_{\set P}h(\vec r)\,d^3r
 \int_{\R^3}\sin^2\left(\gamma \frac{\vec{n}\sca \vec{v}}{\hat v}\right)\,\frac{1}{{\left(\pi \hat{v}^2\right)}^{3/2}}\,
		\exp{\left(-\frac{\vec{v}^2}{\hat v^2} \right)}
\, d^3v
=  \frac{1}{2} \left(1-e^{-\gamma
   ^2}\right)   <\infty\ .
\end{align}
The image of this state under the operator $\op T_V$ is
\begin{align}
\op T_V\, g_\gamma  &= - {\vec v}\sca\n g_\gamma - \frac{e}{m}\n\bar\Phi\sca\n_v g_\gamma\\&
\sim -\frac{e\gamma}{m\hat v}\n\bar\Phi\sca\vec{n}\cos\left(\gamma \frac{\vec{n}\sca \vec{v}}{\hat v}\right)\,\frac{1}{{\left(\pi \hat{v}^2\right)}^{3/4}}\,
		\exp{\left(-\frac{\vec{v}^2}{2 \hat v^2} \right)} \sqrt{\frac{h(\vec r)}{w(\vec r,\vec v)}}. \nonumber
\end{align}
The symbol $\sim$ of the last line indicates that we have displayed only the leading order in $\gamma$. 
Calculating the norm, the leading term obviously diverges for $\gamma\to\infty$, 
\begin{align}
	 \norm{\op T_V g_\gamma}^2 \sim \frac{1}{2}\gamma^2 \left(e^{-\gamma
		 ^2}+1\right) \left(\frac{e}{m\hat v}\n\bar\Phi\sca\vec{n}\right)^2 \to \infty.
\end{align}

Being unbounded, the operator  $\op T_V$ cannot be defined on the full Hilbert space $\set H$ but only on a
dense subset of it,  namely the set of distribution functions $g$ which are differentiable. \linebreak
(In this context, a function $g \in \set H$ is differentiable when its total derivative exists
in the distribution sense, i.e., is also an element of $\set H$. Then also its image under any linear differential operator
is an element of $\set H$, particularly $\op T_V g\in \set H$.) Futhermore, we restrict the domain of the operator to those functions
which obey the specular boundary conditions \eqref{SpecBC} at $\set S$.\linebreak (As differentiability implies continuity, the notion of
boundary conditions is well defined.) 
Altogether, we define the domain $\set D(\op T_V)$ as
\begin{align}
	\set D(\op T_V) = \{ g\in \set{H}\; |\; \hbox{g is differentiable and obeys the conditions \eqref{SpecBC} at $\set S$} \}.
\end{align}

We now consider the scalar product between $\op T_V\, g$ and a state $g^\prime$ which is differentiable, \linebreak
but not necessarily an element of the domain $\set D(\op T_V)$. (That is, it does not necessarily obey the specular boundary condition.)
We first calculate $\langle g^\prime |\op T_V\, g\rangle$:
\begin{align}
	 \langle g^\prime |\op T_V\, g\rangle  =&   \int_{\set P}\int_{\R^3} {g^\prime}^* \left(- {\vec v}\sca\n g - \frac{e}{m}\n\bar\Phi\sca\n_v g    
		+e\n\op\Phi\left\{g\right\}\sca \vec v\right) \, w \, d^3v \,d^3r \\[0.5ex] &
	-  \int_{\set P}\int_{\R^3}e\op\Phi\{{g^\prime}^*\} \left( -{\vec v}\sca\n g - \frac{e}{m}\n\bar\Phi\sca\n_v g
	+e\n\op\Phi\left\{g\right\}\sca \vec v\right)  \, w \, d^3v \,d^3r \nonumber\\[0.5ex]
	 =&   \int_{\set P}\int_{\R^3}\left( {g^\prime}^* \left(- {\vec v}\sca\n g - \frac{e}{m}\n\bar\Phi\sca\n_v g    
	\right)  - e  \vec{v}\sca \left(\nabla \op\Phi\{{g^\prime}^*\} g  \,-   {g^\prime}^* \n\op\Phi\left\{g\right\}\right)\,  \right)w\, d^3v \,d^3r \nonumber.
\end{align}
Next, we now consider $\langle \op T_V\, g| g^\prime\rangle$
and apply some transformations which utilize the properties of $w$ and $\op\Phi$ and contain partial integrations in $\vec{r}$ and $\vec{v}$.
Of course, these operations employ the assumption that $g^\prime$ is differentiable. 
Note that the result is formally identical to the scalar product $-\langle  g|\op T_V\, g^\prime\rangle$, except for a surface integral
over $\set S$:
\begin{align}
	 \langle \op T_V\, g| g^\prime\rangle  =&   \int_{\set P}\int_{\R^3}\left( \left(- {\vec v}\sca\n g^* - \frac{e}{m}\n\bar\Phi\sca\n_v g^*   \right) g'
	-e {\vec v}\sca\bigl( g^* \nabla \op\Phi\{{g'}\}-\nabla\op\Phi\left\{g^*\right\} g^\prime \bigr)\right) w\, d^3v \,d^3r \nonumber\\[0.5ex]
	= & -\int_{\set S}\int_{\R^3} \vec{v}\sca\vec{n} \, g^* g^\prime \, w \, d^3v \,d^2r \label{Formal}\\
	& +  \int_{\set P}\int_{\R^3}\left( g^* \left( {\vec v}\sca\n g^\prime + \frac{e}{m}\n\bar\Phi\sca\n_v g^\prime   \right)
	+e {\vec v}\sca\bigl(\nabla\op\Phi\left\{g^*\right\} g^\prime- g^* \nabla \op\Phi\{{g'}\} \bigr)\right)  w\, d^3v \,d^3r.
	\nonumber
\end{align}
If we take not only $g$, but also $g^\prime$ from the domain $\set D(\op T_V)$, then both distribution functions obey the
specular boundary condition \eqref{SpecBC} and the surface integral term over $\set S$ vanishes. 
This demonstrates that the operator is skew symmetric 
\begin{align}
	\op T_V\;\; \hbox{skew symmetric} \;\;\;\Longleftrightarrow \;\;\; \langle \op T_V\, g| g^\prime\rangle  = - \langle  g|\op T_V\, g^\prime\rangle
	\; \hbox{for all} \; |g\rangle, |g^\prime\rangle \in \set D(\op T_V).
\end{align}

However, a stronger case can be made. Recall the definition of the adjoint operator ${\op T_V}^*$:\linebreak
$\set D({\op T_V}^*)$ contains all $g^\prime \in  \set H$ for which 
$\set D(\op T_V)\!\to\!\set H, g \mapsto \langle \op T_V\, g| g^\prime\rangle$ is a continuous mapping, \linebreak
and ${\op T_V}^* g^\prime$ is the element of $\set H$ which represents that mapping by
 $\langle \op T_V g|  g^\prime\rangle = \langle g|{\op T_V}^* g^\prime\rangle$.\linebreak
Considering expression \eqref{Formal}, it is evident that $\langle \op T_V\, g| g^\prime\rangle$ is a 
bounded functional of $g\!\in\!\set D(\op T_V)$ \linebreak
if and only if the surface integral term over $\set S$ vanishes. This, in turn, is only possible if $g^\prime$\linebreak  obeys the specular boundary condition,
 i.e., is element of $\set D(\op T_V)$.
Thus, $\set D({\op T_V}^*) = \set D(\op T_V)$. \linebreak
(The assumption of differentiability is necessary for the expressions
to be well defined.)\linebreak
Furthermore, when the surface integral over $\set S$ vanishes, $\langle \op T_V\, g| g^\prime\rangle$ is equal to $- \langle  g|\op T_V\, g^\prime\rangle$.
Altogether, this shows that $\op T_V$ is skew self-adjoint:
\begin{align}
	\op T_V\;\; \hbox{skew self-adjoint} \;\;\;\Longleftrightarrow \;\;\; {\op T_V}^* = - \op T_V.
\end{align}

The spectral theorem now states that the spectrum of the operator $\op T_V$ is purely imaginary.
As the operator is evidently real, it is also symmetric with respect to complex conjugation. \linebreak
On physical grounds, we also assume that it is continuous: Consider a point $\vec{r}$ far away from the probe head 
where the equilibrium plasma was assumed to be spatially homogeneous,
$\bar{\Phi}(\vec r) = {\rm const} \equiv 0 $ and $w(\vec r,\vec v) = -\bar{F}^\prime\left(\frac{1}{2} m \vec{v}^2\right)$. Here,  
solutions of the Vlasov equation can be constructed as localized packages of planar waves, centered around a 
solution $(\vec{k},\omega)$ \linebreak
of the corresponding dispersion relation. In contrast to the single plane waves themselves
which are eigenfunctions of $i\omega-\op T_V$ but not elements of $\set H$ 
(as they are not square integrable), the wave packages are elements of the Hilbert space $\set H$
but not eigenfunctions of $i\omega-\op T_V$.
However, they are approximate eigenfunctions; i.e, their images under $i\omega-\op T_V$ are small.
For~any concrete case, one may easily construct a familily of normalized wave packages where the 
images converge to zero. This argument suggests that $i\omega-\op T_V$ has an unbounded inverse; \linebreak
by definition $\omega$ then belongs to the continuous spectrum of $\op T_V$. 
Altogether, we may assume that the Vlasov operator $\op T_V$ has a spectral representation
as displayed in  \eqref{SpectralRepresentationTV}.

\pagebreak

\section{Properties of the collision operator}\label{sec:AppTS}

We study the characteristics of the collision operator $\op T_S$ which is the difference of a multi\-plication operator with $\nu_0 = \nu_0(\vec{r},v)$ and an
integral operator with the kernel $\frac{d\nu}{d\Omega}(\vec{r},v,\vartheta)$. \linebreak
We first keep $\vec{r}$ and $v$ fixed (suppressed in the notation), and focus on the dependence of $\vartheta$. 
An expansion into Legendre polynomials yields, where we have used in the second equation the addition theorem of the spherical
harmonics:
\begin{align}
	\frac{d\nu}{d\Omega}(\vartheta) = \sum_{l=0}^\infty \nu_l \frac{2l+1}{4\pi} P_l(\cos\vartheta) =  \sum_{l=0}^\infty \nu_l\sum_{m=-l}^l Y_{lm}^*(\vec{e}^\prime) Y_{lm}(\vec{e}). 
\end{align}
The completeness relation of the spherical harmonics is
\begin{align}
	\sum_{l=0}^\infty\sum_{m=-l}^l  Y_{lm}(\vec{e})Y_{lm}^*(\vec{e}^\prime) = \delta^{(2)}(\vec{e}-\vec{e}^\prime).
\end{align}
The projection operator $\op P_l$ of a function $g \equiv g(\vec{v})$ on the unit sphere onto the 
angular momentum eigenspace of quantum number $l$ is therefore
\begin{align}
	\op P_l\, g =   \int_\Omega \sum_{m=-l}^l Y_{lm}(\vec{e})Y_{lm}^*(\vec{e}^\prime) \,g(\vec{e}^\prime)\, d\Omega.
\end{align}
Acting on functions on the unit sphere, the collision operator can thus be written 
\begin{align}
	\op T_S\, g  =  -\nu_0\, g+\int_{\Omega} \frac{d\nu}{d\Omega}\, g(\vec{e}^\prime)\,d\Omega^\prime
	= -\sum_{l=1}^\infty \nu_l\,\op P_l\,g.
\end{align}
With respect to the variables $\vec{r}$ and $v$, the operator $\op T_S$ is just a local multiplication operator. 
We can formally write it as an integration operator
\begin{align}
	\op T_S\, g = \int_{\set P}\int_{R^3} K_S(\vec{r},\vec{v},\vec{r}^\prime,\vec{v}^\prime) g(\vec{r}^\prime,\vec{v}^\prime)\, d^3v^\prime d^3 r^\prime,
\end{align}
where the kernel has the form
\begin{align}
	K_S(\vec{r},\vec{v},\vec{r}^\prime,\vec{v}^\prime)  = -\delta^{(3)}(\vec{r}-\vec{r}^\prime) \frac{1}{v^2}\delta(v-v^\prime)
	\sum_{l=1}^\infty \nu_l(\vec{r}^\prime,v^\prime) \, \op P_l.
\end{align}
From this representation, all important characteristics of $\op T_S$ can be deduced: It is \textit{bounded};
the optimal bound is just the absolute maximum of the functions $\nu_l(\vec{r},\vec{v})$ on the phase space. \linebreak
It is obviously \textit{symmetric}, and as bounded, \textit{self-adjoined}. The spectrum is real,
and consists of the negative function values that the $\nu_l(\vec{r},v)$ assume. Depending on the assumptions
made with respect to those functions, the spectrum is either discrete or continuous.

\pagebreak

\acknowledgments
The authors acknowledge support by the Federal Ministry of Education and Research (BMBF) in frame of the project PluTO, and support by the Deutsche 
Forschungsgemeinschaft (DFG) via Graduiertenkolleg GK 1051, Collaborative Research Center TR 87,\linebreak and the Ruhr University Research School. 
Gratitude is expressed
to M.~Lapke, C.~Schulz, R.~Storch, T.~Styrnoll, T.~Mussenbrock, P.~Awakowicz, T.~Musch, and I.~Rolfes, who are or were part of the MRP-Team 
at Ruhr University Bochum.

\pagebreak

\pagebreak


\pagebreak


\pagebreak

\end{document}